\documentclass[aps,prd,twocolumn,preprintnumbers,nofootinbib,amsmath,amssymb]{revtex4-2}
\usepackage{graphicx}
\usepackage{dcolumn}
\usepackage{bm}
\usepackage{amsmath}
\usepackage{braket}
\usepackage[hidelinks]{hyperref}
\usepackage{epstopdf}
\usepackage{placeins}
\usepackage{multirow}
\usepackage{makecell}
\usepackage{cleveref}
\usepackage{xcolor}
\usepackage{dblfloatfix}

\newcommand{\beq}{\begin{eqnarray}}
\newcommand{\eeq}{\end{eqnarray}}
\newcommand{\beqnn}{\begin{eqnarray*}}
\newcommand{\eeqnn}{\end{eqnarray*}}

\usepackage[font=small, labelfont=bf, textfont={small,it}]{caption}
\def\spose#1{\hbox to 0pt{#1\hss}}
\def\ltapprox{\mathrel{\spose{\lower 3pt\hbox{$\mathchar"218$}}
	\raise 2.0pt\hbox{$\mathchar"13C$}}}

\graphicspath{{figs/}}

\begin{document}

\title{The Roberge-Weiss transition for QCD in a magnetic background}

\author{Massimo D'Elia}
\email{massimo.delia@unipi.it}
\affiliation{Dipartimento di Fisica dell'Universit\`a di Pisa \& \\ INFN Sezione di Pisa, Largo Pontecorvo 3, I-56127 Pisa, Italy}

\author{Lorenzo Maio}
\email{lorenzo.maio@roma2.infn.it}
\affiliation{Dipartimento di Fisica, Universit\`a di Roma “Tor Vergata” \& \\ INFN Sezione di Roma Tor Vergata, Via
della Ricerca Scientifica 1, I-00133 Rome, Italy}

\author{Kevin Zambello}
\email{kevin.zambello@pi.infn.it}
\affiliation{Dipartimento di Fisica dell'Universit\`a di Pisa \& \\ INFN Sezione di Pisa, Largo Pontecorvo 3, I-56127 Pisa, Italy}

\author{Giuseppe Zanichelli}
\affiliation{Dipartimento di Fisica dell'Universit\`a di Pisa \& \\ INFN Sezione di Pisa, Largo Pontecorvo 3, I-56127 Pisa, Italy}

\date{\today}

\begin{abstract}
  We investigate how a magnetic background field influences the location and the nature
  of the Roberge-Weiss (RW) finite temperature transition for $N_f = 2+1$ QCD with physical quark masses.
  To that purpose, we perform numerical simulations of the finite temperature theory, discretized through
  stout staggered quarks and the tree-level improved Symanzik pure gauge action, considering
  two different values of the Euclidean temporal extent in lattice units, $N_t = 6, 8$.
  The RW transition temperature $T_{RW}$ decreases with $eB$,
  in particular it follows closely the behavior 
  of the pseudo-critical QCD crossover temperature $T_{pc}$, so that
  $T_{RW} (eB) - T_{pc}(eB)$ is practically constant, within errors, for magnetic fields up to
  $eB \sim 1$~GeV$^2$; consistent results are found from the drop of the chiral condensate, which signals chiral
  symmetry restoration, leading also to the phenomenon of inverse magnetic catalysis above the transition.
  Moreover, we find that the magnetic field turns the RW transition
  from second order to first order, with a tri-critical magnetic field in-between
  1 and 2.4~GeV$^2$, i.e. for magnetic fields substantially lower than those
  for which the standard QCD transition turns to first order.
  \end{abstract}

\maketitle

\section{Introduction}

The phase diagram of Quantum Chromodynamics (QCD) is still the object of extensive investigation.
In the purely finite temperature theory, QCD is known to undergo a crossover
from a hadronic phase to a QGP phase, at a pseudocritical temperature $T_{pc} \simeq 155-160$~MeV for physical quark
masses~\cite{Aoki:2006we,Aoki:2006br,Borsanyi:2010bp,Bazavov:2011nk,Bhattacharya:2014ara}.
A real phase transition is present only 
in the limit of infinite or zero quark masses, where exact global symmetries, 
respectively center and chiral symmetry, appear, with an associated spontaneous breaking
which characterizes the phase transition.

The introduction of external parameters can lead to interesting and physically relevant extensions
of the phase diagram. A notable example is the baryon chemical potential, which enters the description
of finite baryon density systems, like those reproduced in heavy-ion collisions or in compact astrophysical objects,
and is expected
to strengthen the QCD crossover till it becomes a real transition at a critical endpoint
in the $T - \mu_B$ plane, after which there should be a first order transition in the low-$T$ - high-$\mu_B$ region of the phase diagram. 

Such critical endpoint is the object of active theoretical and experimental searches \cite{Fodor:2001pe,Fu:2019hdw,Braguta:2019yci,Gao:2020fbl,Gunkel:2021oya,Hippert:2023bel,Clarke:2024ugt,Lysenko:2024hqp,Shah:2024img,Adam:2025pii,Borsanyi:2025dyp};
however, unfortunately, the study of the $T - \mu_B$ phase diagram by means of lattice QCD simulations is hindered
by the presence of the sign problem. On the other hand, considering an imaginary (instead of real) baryon chemical potential 
is free of the sign problem and, apart from being a possible way of obtaining some pieces of information regarding small
real $\mu_B$ physics via analytic continuation~\cite{Alford:1998sd,Lombardo:1999cz,deForcrand:2002hgr,DElia:2002tig},
it represents a (theoretically) appealing extension of the phase diagram per se.

In particular, for certain values of the imaginary 
baryon chemical potential, $\mu_B/T = i \pi (2 k + 1)$ with $k$ a relative integer, the theory presents an exact
remnant of the global center symmetry, which is expected to break at a critical temperature $T_{RW}$ (Roberge-Weiss transition)~\cite{Roberge:1986mm}.
$T_{RW}$ represents, in some sense, the counterpart of the deconfining transition of pure Yang-Mills and appears to be continuously connected
to the QCD crossover
at $\mu_B = 0$. According to the residual center symmetry, the transition is expected to belong to the 3D-Ising universality class
if it is second order, even if a first order transition is not excluded and has indeed been observed for certain ranges
of quark masses~\cite{DElia:2009bzj,deForcrand:2010he,Bonati:2010gi,Cea:2012ev,Wu:2013bfa,Philipsen:2014rpa,Wu:2014lsa,Nagata:2014fra,Kashiwa:2016vrl,Makiyama:2015uwa,Czaban:2015sas,Philipsen:2019ouy}.
Continuum extrapolated results for $N_f = 2+1$ QCD at the physical point have confirmed the presence of a 
second order transition at  a temperature $T_{RW} \simeq 210$~MeV~\cite{Bonati:2016pwz,Dimopoulos:2021vrk},
and strong hints exist that at the same
temperature chiral symmetry is restored as well~\cite{Bonati:2018fvg,Cuteri:2022vwk}.

A different and phenomenologically relevant extension of the phase diagram is represented by the introduction of a magnetic background field,
which couples to QCD through the quark electric charges. Despite electromagnetic interactions being in general a small correction
to strong interactions, the effects can be relevant when $e B \gtrsim \Lambda_{QCD}^2$, thus affecting the description of non-central heavy-ion collisions, of magnetars, and possibly of the early stages of the Universe. Differently from the case of a baryon chemical potential, the introduction of a magnetic background field
leaves the path integral measure real and positive, so that extensive lattice simulations are possible. Indeed, many
properties of strong interactions in a magnetic background have been clarified by lattice QCD simulations,
including in particular those regarding the $T - eB$ phase diagram (for a recent review, see Ref.~\cite{Endrodi:2024cqn}). 

After some initial contradictory results, mostly due to the presence of large discretization effects~\cite{DElia:2010abb},
numerical simulations have demonstrated a consistent decreasing behaviour
of the pseudocritical temperature $T_c$ as a function of $eB$~\cite{Bali:2011qj},
which is found independently of the quark mass spectrum~\cite{DElia:2018xwo,Endrodi:2019zrl}
and is similar to the corresponding behaviour of $T_c$ as a function of $\mu_B$.
Moreover the magnetic background field makes the change
of thermodynamical observable around $T_c$ steeper~\cite{DElia:2010abb,Endrodi:2015oba,Ding:2020inp}, pointing to the possible appearance of a real phase transition
for large enough magnetic
fields. Recent studies~\cite{DElia:2021yvk} have proved that indeed, for $N_f = 2+1$ QCD with physical quark masses,
the transition becomes first order
at large $eB$, with a critical endpoint located approximately in the range
$[65 - 95]~\textrm{MeV} \times [4-9]~\textrm{GeV}^2$ in the $T - e B$ plane.

Many other non-trivial properties have been discussed in the literature:
Refs.~\cite{Astrakhantsev:2019zkr,Almirante:2024lqn} investigated the electrical
conductivity of the quark-gluon plasma in the presence of a background magnetic field, finding evidence
for an enhancement of the conductivity along the magnetic field and a supression transverse to it;
the effects on the fluctuations of conserved charges have been investigated
in Ref.~\cite{Ding:2023bft}, while
Refs.~\cite{Bonati:2014ksa,Bonati:2016kxj,Bonati:2017uvz,DElia:2021tfb} investigated the effect of
magnetic fields on the confining properties of QCD.

The presence of a critical endpoint at which the QCD crossover turns into first order makes the 
$T - e B$ phase diagram close to the structure expected in the $T - \mu_B$ one.
Therefore, the investigation of the QCD phase diagram, considering at the same time a magnetic background field
and a finite baryon chemical potential $\mu_B$ looks interesting. In particular, one would like to understand if the
similarities reflect in a connection between the phase structures when one considers the three dimensional
$T - \mu_B - eB$ phase diagram.
For instance, if the appearance of a first order transition is related to similar physical effects in the two cases,
the critical endpoint should move to lower values of $eB$ as a
chemical potential $\mu_B$ is switched on, since the effects of the two external parameters would add to each other.

As usual, being unable to combine and directly simulate $eB$ and real $\mu_B$, because of the sign problem,
we will look for our keys under the street lamp of an imaginary chemical potential, 
hoping to learn at least something interesting, and useful to some extent, from that.
The investigation of finite temperature QCD under the combined effect of magnetic fields and imaginary chemical potentials
has been already considered in the literature~\cite{Braguta:2019yci}.
In this study we focus directly on the Roberge-Weiss phase transition and
on how it is influenced by the presence of a magnetic background field.
There are two main questions which are worth addressing.
Is the dependence of $T_{RW}$ on $eB$ similar to what already observed for $T_c(eB)$? Is there any critical value
of the magnetic field, at which the second order Roberge-Weiss transition observed for $eB = 0$ turns into
first order?

We adopt the same discretization adopted in
our previous studies,
namely $N_f =2+1$ QCD with physical quark masses and a stout improved staggered discretization, with various values of $eB$
and two different temporal extents, $N_t = 6,8$, in order to provide a first assessment of finite cut-off effects.
The paper is organized as follows. In Section~\ref{sec:setup} we illustrate the adopted discretization and our numerical setup.
In Section~\ref{sec:results} we show and discuss our results. Finally, in Section~\ref{sec:conclusions} we draw our conclusions.

\section{Numerical Setup}

\label{sec:setup}

We consider a discretization of $N_f = 2+1$ QCD
based on the tree-level improved Symanzik pure gauge action~\cite{weisz,curci}
and on stout rooted staggered fermions~\cite{kogut-susskind,morningstar},
i.e.~on the following partition function
\begin{equation}
  Z=\int{[DU]}\,e^{-S_{YM}}\prod_{f=u,d,s} \det{(M_{st}^f)}^\frac{1}{4},
\end{equation}
where $[DU]$ is the Haar measure for gauge links,
$f$ is the flavor index, while the fermion matrix and the gauge action
are respectively
\beq
  {M^f_{st\ }}_{ij} &=& \hat m_f\delta_{ij}+\sum_{\nu=1}^{4}\frac{\eta_{i;\nu}}{2}
\left(U^{(2)}_{i;\nu}\delta_{i\, j-\hat{\nu}} - U^{(2)\dagger}_{i-\hat{\nu};\nu}\delta_{i\,j+\hat{\nu}}\right) \nonumber \\
\label{dirac_operator}
  S_{YM} &=& -\frac{\beta}{3} \, \sum_{\substack{i \\ \mu\ne\nu}}\left(\frac{5}{6}W^{1\times1}_{i,\mu\nu}-\frac{1}{12}W^{1\times2}_{i,\mu\nu}\right)
\eeq
with periodic (antiperiodic) boundary conditions in the Euclidean temporal direction
for bosonic (fermionic) fields, in order to reproduce a thermal system.
There, $i,j$ and $\mu,\nu$ stand respectively for lattice sites and directions,
$\beta$ is the inverse gauge coupling, $a$ is the lattice spacing, while
$\hat m_f = a m_f$, $f = u,d,s$, are the dimensionless bare quark masses.
The $\eta_{i;\nu}$ are the staggered quark phases,
$U^{(2)}_{i;\nu}$ is the two-times stout smeared link
(with isotropic smearing parameter $\rho=0.15$), while
$W^{1\times 1,2}_{i,\mu\nu}$ are the real parts of the trace
of the link products along the $1\times1$ and $1\times2$ rectangular closed paths,
respectively.

Bare quark masses and the gauge coupling values have been tuned in order to move
along a line of constant physics (LCP), which reproduces
experimental results for hadronic observables,
based on the determinations reported in Refs.~\cite{Aoki:2009sc,Borsanyi:2010cj,Borsanyi:2013bia}.
Contrary to the strategy adopted in Ref.~\cite{DElia:2021yvk}, in this 
study we consider two fixed values of the temporal extension in lattice units, 
$N_t = 6,8$, and for each extension we tune the physical temperature
by changing $\beta$, hence the lattice spacing, along the LCP:
this strategy has some drawbacks, including a variation of $eB$ when
$T$ is changed, however it allows for a better fine tuning of the 
temperature, which is useful in this context because of the extended 
set of explored magnetic fields.

\subsection{Finite B and finite imaginary chemical potential on a lattice}
In the lattice approach, both the presence of a
magnetic background field and that of an imaginary
chemical potential can be translated in the introduction
of additional $U(1)$ phases to the elementary parallel transporters
\begin{equation}
  U^{(2)}_{i;\mu} \to u_{i;\mu}^f U^{(2)}_{i;\mu} \, .
\end{equation}
Such phases are considered as constant external parameters, since no functional integration is performed
over them, and take different values for each flavor, depending
on the flavour electric charge and chemical potential.
The magnetic background determines the $U(1)$ phases of spatial links, while the chemical potentials
those of temporal links, so that they are decoupled from each other at least at the level
of the bare Lagrangian, i.e.~apart from effects which emerge at a dynamical level
and that we are going to explore by numerical simulations. That means, for instance,
that the exact $\mathbb{Z}_2$ global symmetry emerging along the Roberge-Weiss lines, and the definition of the
associated order parameters, are not affected by the presence of the magnetic background.

In the following, we will consider a uniform 
magnetic background directed along the $\hat{z}$ direction.
A possible discretization on a periodic toroidal lattice,
corresponding to the gauge choice $A_x= A_z= 0,\ A_y(x) = Bx$,
is the following
\begin{equation}
  u_{i;y}^f=e^{i a^2 q_f B \,i_x}, \qquad {u_{i;x}^f\vert}_{i_x=L_x}=e^{-ia^2q_fL_xBi_y},
\end{equation}
with all other $U(1)$ link  variables set to one.
 $L_i$ is the number of lattice sites
along direction $i$, and the boundary values for the $x$-links 
are added to make the magnetic field smooth across the
$x$-boundary~\cite{tHooft:1979rtg,wiese,review}.
Actually, smoothness is guaranteed but for a single
plaquette, which is pierced by an additional Dirac string taking charge 
of the zero magnetic flux condition across the lattice torus; invisibility
of such string leads to a quantization condition of $B$,
which is dominated by the smallest quark charge $q_f = e/3$, leading to 
\begin{equation}
  q_f B=\frac{ 2 \pi b_z}{a^2L_xL_y} \implies
  e B=\frac{ 6 \pi b_z}{a^2L_xL_y}, \qquad b_z \in \mathbb{Z} \, .
\label{def_B}
\end{equation}
In a finite temperature setup, where $T = 1 / (N_t a)$, that implies
\begin{equation}
  \frac{e B}{T^2} = \frac{ 6 \pi b_z N_t^2}{L_xL_y} \, ,
\label{def_B_2}
\end{equation}
meaning that, adopting our strategy for which $T$ is tuned by changing $a$
along the LCP at fixed $N_t$, the $T - eB$ phase diagram will be explored moving
along lines with fixed $b_z$ (in order to keep the quantization condition
stable against small $T$ variations),~i.e., $eB$ will change proportionally to $T^2$.
This is only a minor drawback, since we are just interested in determining the transition points
along these lines, for which both $T$ and $eB$ will be well defined anyway.
Moreover, also the study of the critical behavior will be unaffected, since both
$T$ and $eB$ preserve the $\mathbb{Z}_2$ symmetry which drives the transition.
The magnetic field couples to dynamical quarks through the gauge
invariant $U(1)$ phase factors
that quarks pick up going through
closed loops on the lattice. In order to keep discretization
effects under control, the elementary phase factor corresponding
to the smallest non-trivial loop, a plaquette in the $xy$ plane,
must be small enough: this is guaranteed if the elementary
flux $e B a^2 \ll 2 \pi$, a condition which in a finite $T$ setup
can also be expressed as $e B/T^2 \ll 2 \pi N_t^2$.
In the following, the lowest explored temperatures will be approached for the largest
explored magnetic fields and will correspond to around
$150$~MeV, which translates the condition into $eB \ll 0.14~N_t^2$~GeV$^2$. Such cutoff
corresponds to around 5~GeV$^2$ for $N_t = 6$ and 9~GeV$^2$ for $N_t = 8$:
since our largest values of $eB$ are above the GeV$^2$ scale, exploring
at least two values of $N_t$ is mandatory in order to see if significant differences
emerge in the two cases.
\\

Chemical potentials affects only the temporal link variables.
In particular, for a purely imaginary chemical potential one has
\begin{equation}
  u_{i;t}^f=e^{i a {\rm Im} (\mu_f) } \, .
\end{equation}
Such constant temporal phases can also be viewed as a global
rotation of fermionic boundary
conditions in the temporal direction,
by an angle $\theta_f = a N_t {\rm Im} (\mu_f) = {\rm Im} (\mu_f)/T$.
In the presence of a purely baryonic chemical potential,
i.e.~when the quark chemical potentials are set equal for all flavors,
$\mu_u = \mu_d = \mu_s \equiv \mu_q = \mu_B/3$, one expects
a $2 \pi$-periodicity in $\theta_q= {\rm Im} (\mu_q)/T$.
However, the actual periodicity is $2 \pi/N_c$,
since this rotation of fermionic boundary conditions can be exactly 
reabsorbed by a center transformation on the gauge fields, which
leaves the rest of the path integral (pure gauge term and measure) unchanged~\cite{Roberge:1986mm}.
Numerical simulations show that the $2 \pi/N_c$ periodicity in $\theta_q$
is smoothly realized at low temperatures~\cite{deForcrand:2002hgr,DElia:2002tig}, while it is enforced
by first order phase transitions at high-$T$, which can be understood as follows.

It is known that in pure gauge theories
the Polyakov loop $L$, i.e.~the trace of the temporal Wilson line normalized by $N_c$,
is an order parameter for the spontaneous breaking of center symmetry, and
develops a non-zero expectation value at high $T$. With dynamical fermions,
the direct coupling of the fermion determinant to the Polyakov loop
breaks center symmetry explicitly, however in the presence of an
imaginary baryon chemical potential $L$ enters the fermionic
determinant expansion multiplied by $\exp (i \theta_q)$, hence it is the value of
$\theta_q$ which selects the true vacuum among the $N_c$ different minima of the
Polyakov loop effective potential, which are related to each other by center
transformations. Therefore, phase transitions occur when $\theta_q$ crosses the
boundary between two center sectors, i.e.  for $\theta_q = (2 k +
1)\pi/N_c$ and $k$ integer, where $\langle L \rangle$ jumps from one center sector to the
other~\cite{Roberge:1986mm}.
The $T$-$\theta_q$ phase diagram then consists of a periodic
repetition of first order lines (RW lines) in the high-$T$ regime, which
disappear at low $T$. Therefore they have an endpoint at some temperature
$T_{\rm RW}$, where an exact $\mathbb{Z}_2$ symmetry breaks spontaneously: this
is exactly the Roberge-Weiss transition which is the subject of the present
investigation.

The phase of $L$ can serve as a possible order parameter
for the spontaneous breaking of the residual $\mathbb{Z}_2$ center symmetry
along the RW-lines; an alternative order parameter would be the imaginary part
of any of the quark number densities.
It is convenient to choose in particular the RW-line corresponding to 
to $\theta_q = \pi$, since in this case one can take as an order parameter
the imaginary part of the Polyakov loop.
The order parameter susceptibility is then defined as
\begin{equation}\label{suscdef}
\chi_L \equiv N_t N_s^3\ (\langle ({\rm Im}(L))^2 \rangle - \langle
|{\rm Im}(L)| \rangle^2) \, ,
\end{equation}
where $N_s$ ($N_t$) is the spatial (temporal) size in lattice units.
{On a finite lattice, the order parameter and other relevant thermodynamical
  quantities present an analytic behavior around the transition, and their critical
  behavior is revealed by their finite size scaling (FSS). In particular,}
the
susceptibility $\chi_L$ is expected to scale, around the RW-transition at
fixed $N_t$ and $\theta_q$, as 
\begin{equation}\label{fss}
\chi_L = N_s^{\gamma/\nu}\ \phi (\tau N_s^{1/\nu}) \, , 
\end{equation}
where $\tau$ is the reduced temperature, $\tau = (T - T_{\rm RW})/T_{\rm RW}$,
while the critical indices characterize the order of the transition and, in case of a continuous
transition, its universality class.

The effective theory associated with the spontaneous breaking of the $\mathbb{Z}_2$
symmetry is the three dimensional Ising model,
so the transition can be either first order or second order in the
3D Ising universality class, with a tricritical behavior separating
the two possible regimes. Therefore,
apart from the unlikely case of being exactly on the tricritical point,
the ultimate large volume FSS behavior should be
either first order or Ising $3d$~\cite{LawSarb, Bonati:2010gi,
Bonati:2010ce,Bonati:2013ota}. The critical indices that will be used in the following analysis are
reported for convenience in Table~\ref{tab:critexp}.

\begin{table}[bt!]
\begin{tabular}{|c|c|c|c|c|}
\hline                & $\nu$     & $\gamma$    & $\gamma/\nu$ & $1/\nu$\\
\hline $1^{st}$ Order & 1/3       & 1           & 3            & 3\\
\hline $3D$ Ising     & 0.6301(4) & $1.2372(5)$ & $\sim 1.963$ & $\sim 1.587$ \\
\hline Tricritical & 1/2       & 1           & 2            & 2\\
\hline
\end{tabular}
\caption{The critical exponents relevant for this study
  (see, e.g., Refs.~\cite{Pelissetto:2000ek,1995JPhA...28.6289B}).}\label{tab:critexp}
\end{table}

\section{Numerical Results}

\label{sec:results}

We have performed numerical simulations on lattices with two different
temporal extensions, $N_t = 6$ and $N_t = 8$, to the purpose of assessing
finite cut-off effects, and various spatial extensions in some cases,
to the purpose of assessing the critical behaviour at the transition.
For $N_t = 6$ we have considered $7$ different values of the magnetic field,
$eB = 0.2, 0.4, 0.6, 1.0, 1.2, 1.6$ and 2.4~GeV$^2$. In all cases,
the value of the magnetic field is that computed at the critical temperature,
and the value of $b_z$ in Eq.~(\ref{def_B}) has been tuned when changing the
spatial size so as to maintain $eB$ constant at the percent level.

A FSS analysis has been performed only for $eB = 1.0$ and 2.4~GeV$^2$; in particular, as we will
show in the following, we have found indications for a second order and for a first
order transition, respectively. Only in those cases we have repeated the analysis
also on $N_t = 8$ lattices, in order to see if these findings are stable
as one approaches the continuum limit. In the following we will start showing
results obtained for $N_t = 6$. Then we will illustrate those for $N_t = 8$ and,
finally, we will discuss
a sketch of the Roberge-Weiss critical line in the $T - eB$ plane and compare
it with the analogous behavior of the standard QCD thermal transition.

\begin{figure}[t!]
  \centering
  \includegraphics[width=1.0\linewidth, clip]{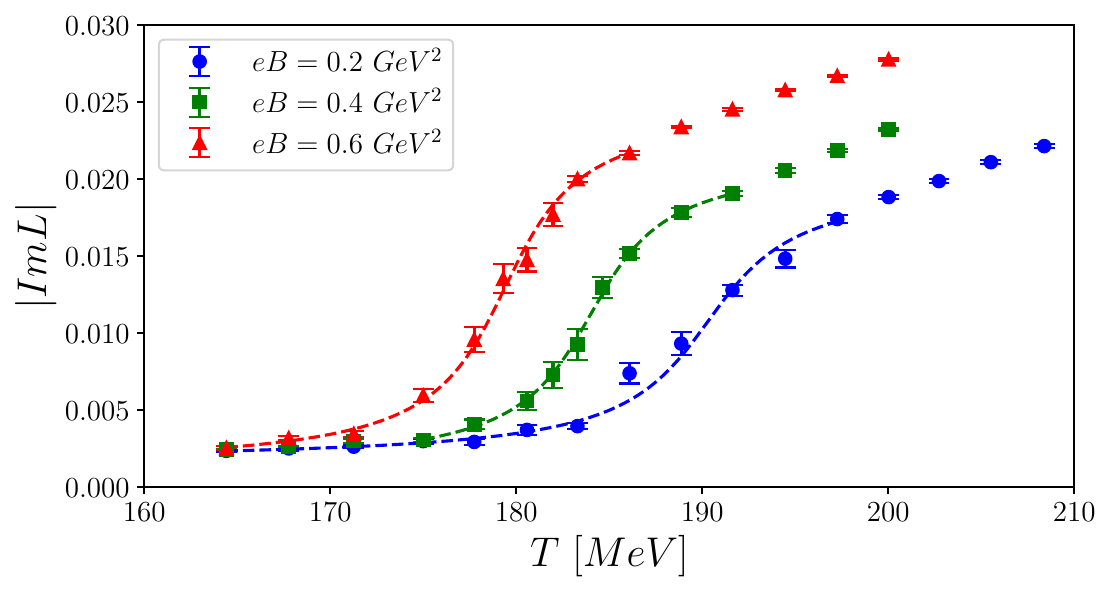}
  \caption{Average value of the modulus of the imaginary part of the Polyakov
  loop as a function of $T$ for $eB = 0.2,~ 0.4$ and $0.6$~GeV$^2$ on $N_t=6$ lattices.}
  \label{fig:smallB_L}
\end{figure}

\begin{figure}[t!]
  \centering
  \includegraphics[width=1.0\linewidth, clip]{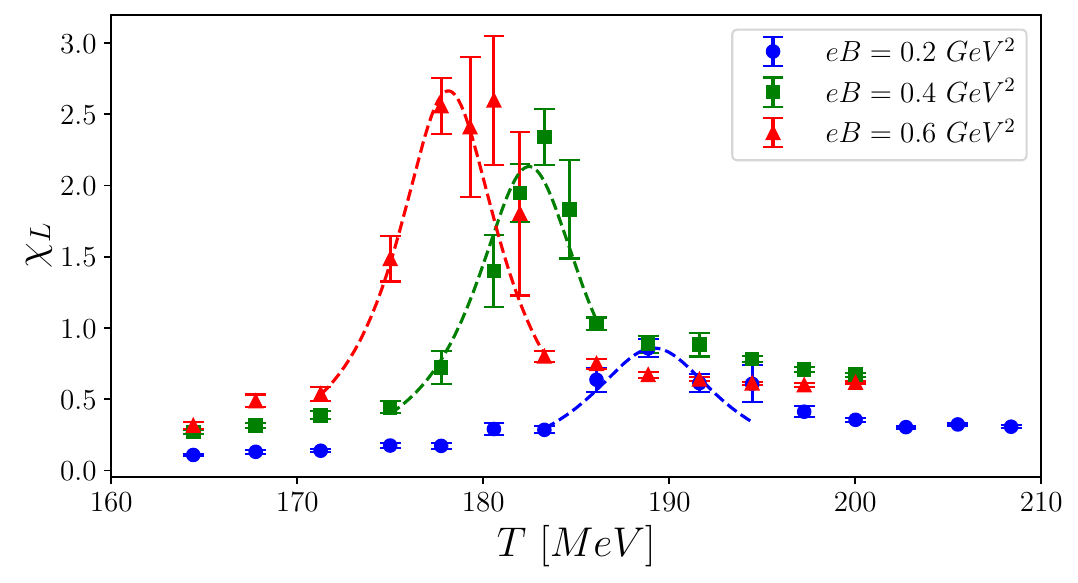}
  \caption{Susceptibility of the imaginary part of the Polyakov
  loop as a function of $T$ for $eB = 0.2,~ 0.4$ and $0.6$~GeV$^2$ on $N_t=6$ lattices.}
  \label{fig:smallB_chiL}
\end{figure}

\begin{figure}[t!]
  \centering
  \includegraphics[width=1.0\linewidth, clip]{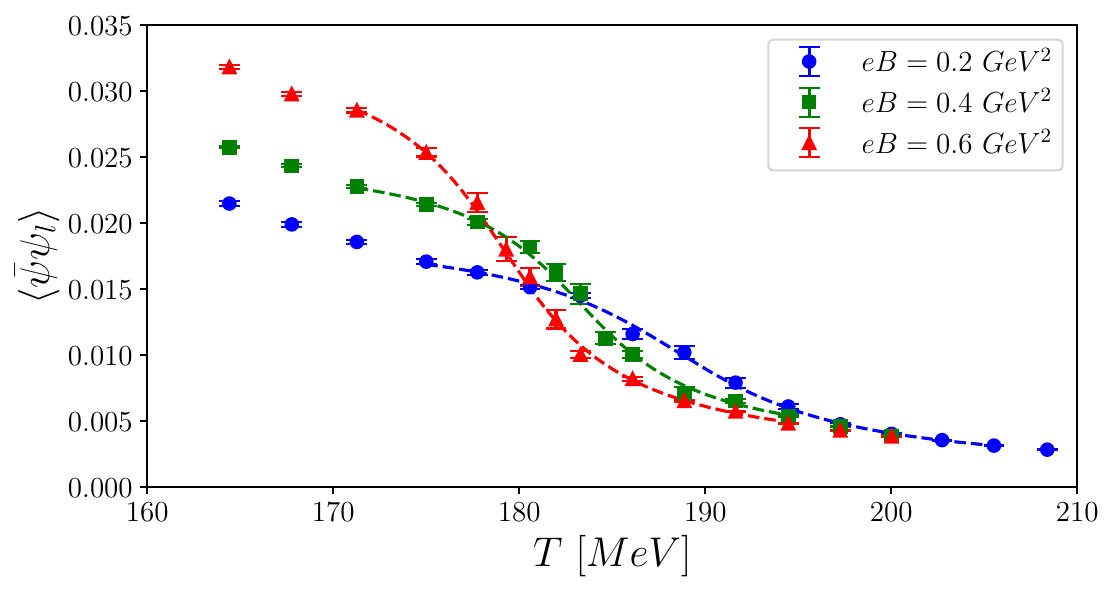}
  \caption{Unrenormalized light chiral condensate as a function of $T$
    for $eB = 0.2,~ 0.4$ and $0.6$~GeV$^2$ on $N_t=6$ lattices.}
  \label{fig:smallB_FF}
\end{figure}

\subsection{Results for $N_t = 6$}

We performed an initial series of runs at $eB = 0.2,~ 0.4$ and $0.6$~GeV$^2$ on $N_t=6$ lattices,
considering only two spatial sizes, $N_s = 18,~24$.
In Figs.~\ref{fig:smallB_L} and~\ref{fig:smallB_chiL} we show results obtained for the
order parameter (imaginary part of the Polyakov loop) and its susceptibility.
The Roberge-Weiss temperature, determined either from the
inflection point of the order parameter through an arctan fit,
or from the peak of the susceptibility through a Lorentzian fit,
is found to decrease monotonically with  an increasing magnetic field.
Even if we have not performed a careful FSS analysis in these cases,
the determination of the critical temperature is not significantly affected by $N_s$:
for instance at $eB = 0.6$~GeV$^2$ we obtain $T_{RW} = 180.30(43)$~MeV
for $N_s=18$ and $T_{RW} = 178.99(59)$~MeV for $N_s=24$.

We have also performed an analysis of the unrenormalized light chiral condensate around the transition,
which is shown in Fig.~\ref{fig:smallB_FF}. The drop of the condensate, which signals the restoration
of chiral symmetry, moves with $eB$ consistently with what observed for $T_{RW}$. Morover,
one clearly observes the phenomenon of magnetic catalysis at low temperatures, i.e.~an increase of the
condensate with $eB$ at a given temperature, which turns into inverse magnetic catalysis around or
above the transition, as a consequence of the drop of the transition temperature.
We notice that, even if this part of our analysis is still based on unrenormalized quantities,
the subtraction of data at zero temperature and magnetic field, which is needed to properly renormalize
the condensate, is not expected to affect the observations above in a significant way.

\begin{figure}[t!]
  \centering
  \includegraphics[width=1.0\linewidth, clip]{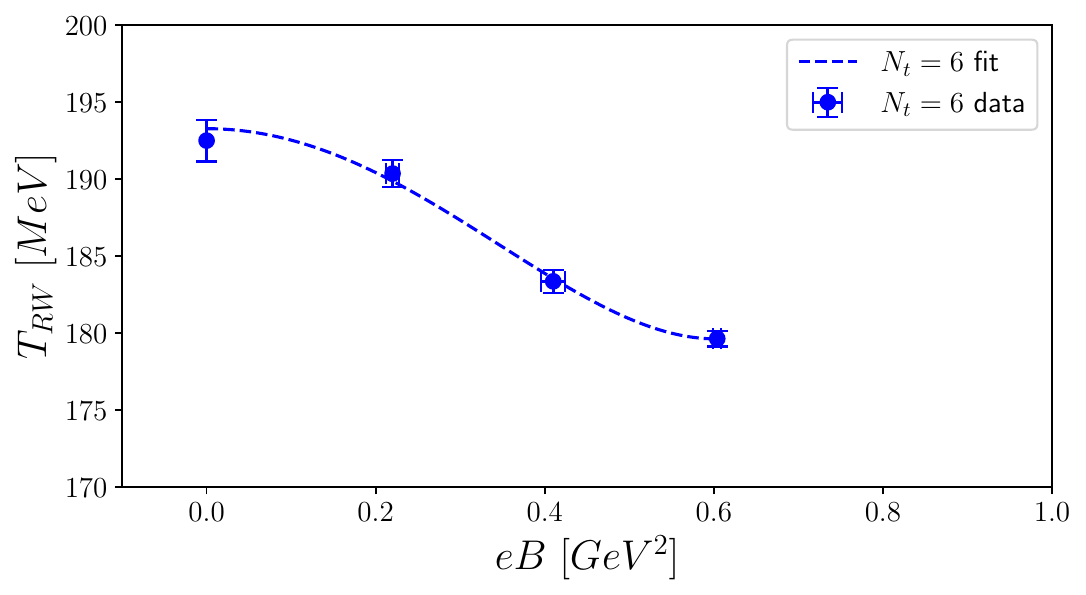}
  \caption{Roberge-Weiss transition line for small eB}
  \label{fig:critline_loweb}
\end{figure}

Numerical results for $T_{RW}$ obtained from the susceptibility on $N_s = 24$ lattices
are illustrated in Fig.~\ref{fig:critline_loweb}, where we also show the result
at $eB = 0$~GeV$^2$ from a previous work. 
The dashed line is the result of a best fit using the rational function ansatz
$T_{RW}(eB) = T_{RW}(0) \frac{1 + a(eB)^2}{1 + b(eB)^2}$, which is the same ansatz employed
in Ref. \cite{Endrodi:2015oba} to parametrize the chiral transition line on the $(\mu_q=0, T, eB)$ plane. The ansatz describes well our data, at least for these relatively low magnetic fields. We notice that the critical line changes curvature around $eB \sim 0.6$~GeV$^2$, close to the magnetic flipping point where
a qualitative change in the behavior of the physical curvature
of the chiral transition at non-zero $\mu_B$ was observed in Ref.~\cite{Braguta:2019yci}.

\begin{figure}[t!]
  \centering
  \includegraphics[width=1.0\linewidth, clip]{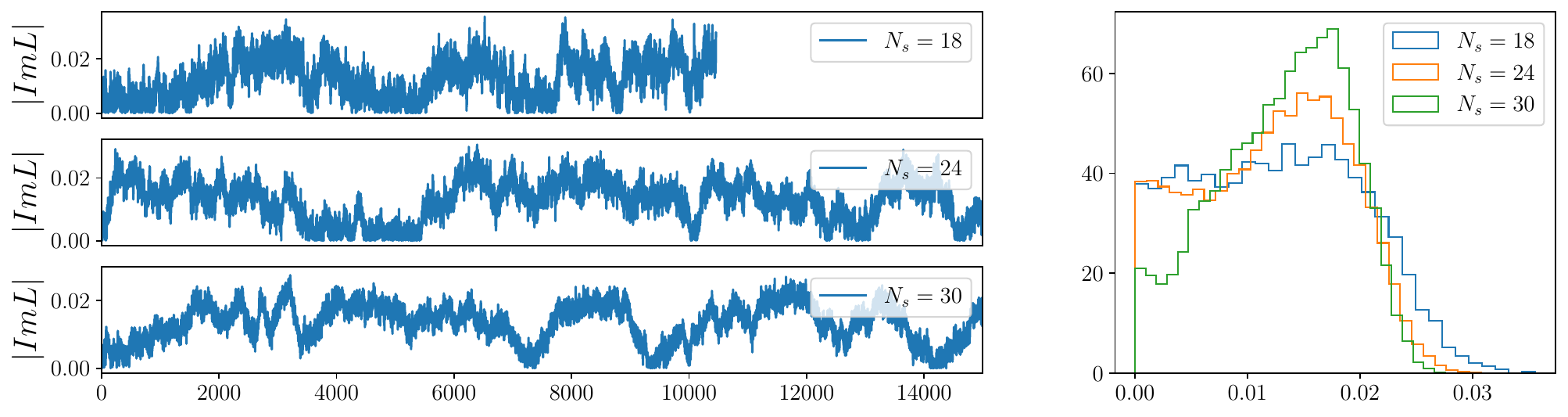}
  \caption{MC histories and histogram: $eB=1.0$ $GeV^2$, $N_t=6$.}
  \label{fig:1p0GeV2_Nt6_hist}
\end{figure}

\begin{figure}[t!]
  \centering
  \includegraphics[width=1.0\linewidth, clip]{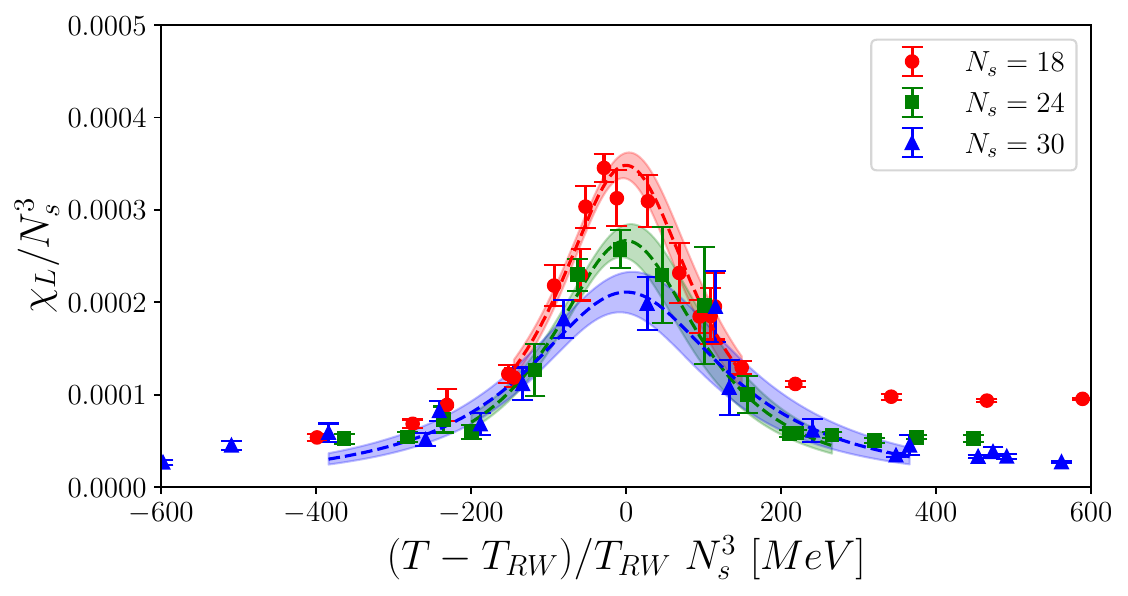}
  \includegraphics[width=1.0\linewidth, clip]{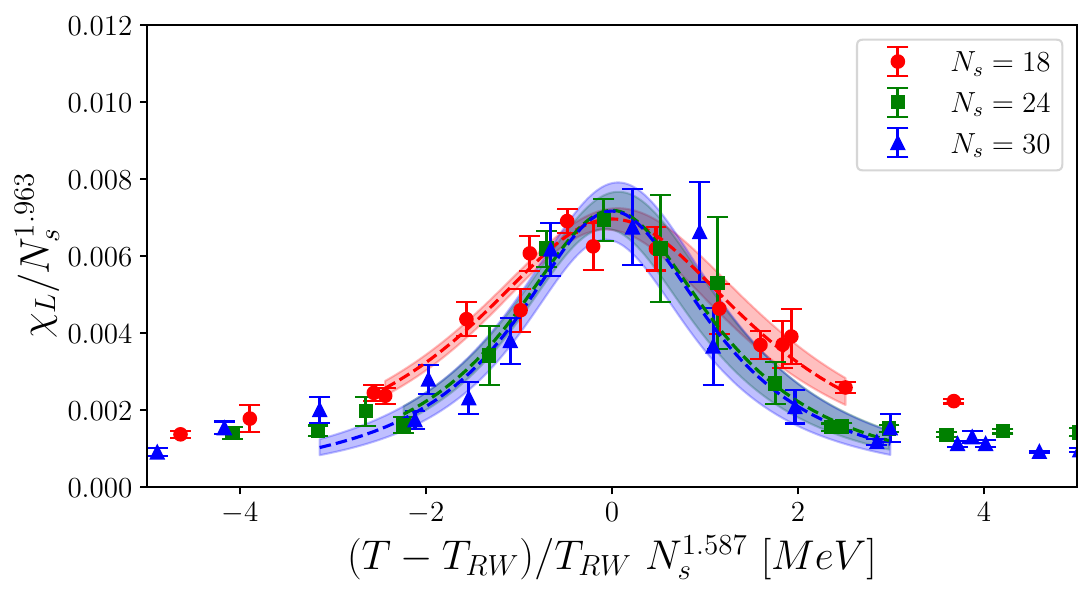}
  \includegraphics[width=1.0\linewidth, clip]{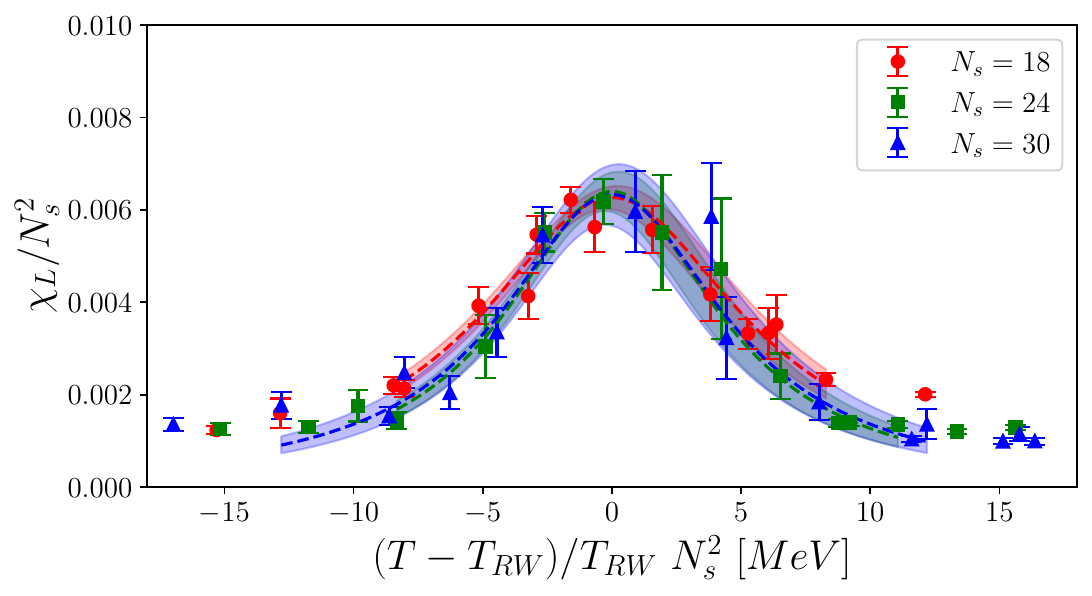}
  \caption{Collapse plots of the order parameter susceptibility for $eB=1.0$ $GeV^2$, $N_t=6$, according to first order (top), second order $Z_2$ (center), tricritical (bottom).}
  \label{fig:1p0GeV2_Nt6_collapseplots}
\end{figure}

\begin{figure}[t!]
  \centering
  \includegraphics[width=1.0\linewidth, clip]{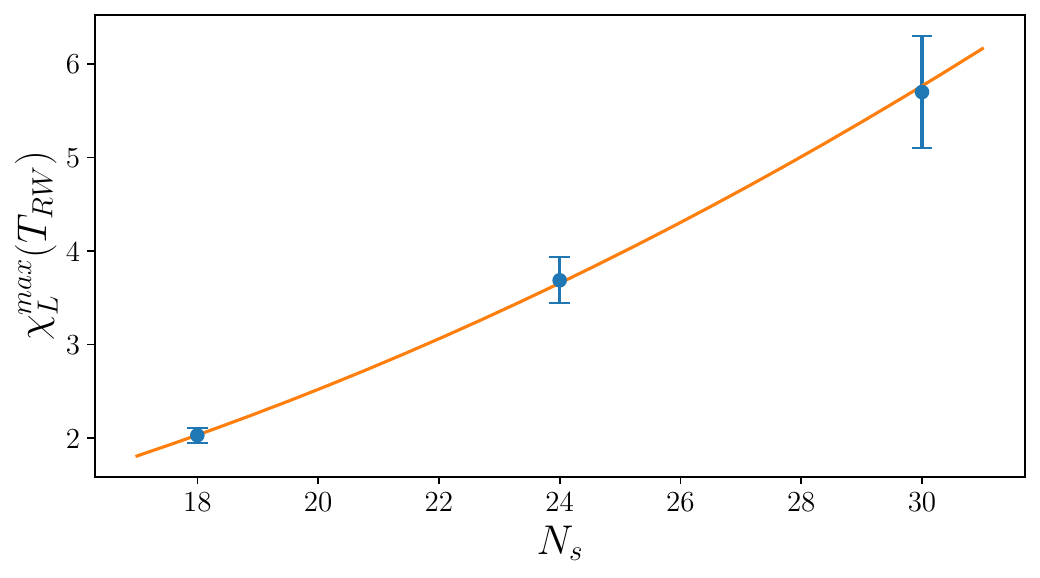}
  \caption{Fit for $\chi_L^{max}$: $eB=1.0$ $GeV^2$, $N_t=6$.}
  \label{fig:1p0GeV2_Nt6_suscfit}
\end{figure}

We conducted additional runs in the strong magnetic field regime, in particular
for $eB = 1.0$ and $2.4$~GeV$^2$. Simulations for
$eB = 1.0$~GeV$^2$ were performed for three different spatial extensions,
$N_s=18,24,30$. In Fig.~\ref{fig:1p0GeV2_Nt6_hist} we report
some Monte Carlo histories (left panel) and histograms (right panel)
of the order parameter obtained for the three volumes around the transition temperature.
The absence of clear double peaked structures in the histograms suggests that the transition
is still continuous at this value of the magnetic field, as it is at $eB = 0$.
This is confirmed by a FSS analysis. The susceptibility is expected to scale as
in Eq.~(\ref{fss}), therefore, when plotting
$\chi_L  / N_s^{\frac{\gamma}{\nu}}$ as a function of $\tau N_s^\frac{1}{\nu}$,
the plots for the different lattice sizes are expected to collapse onto each other
when the correct critical indices are used. Moreover, Eq.~(\ref{fss}) implies that
the heights of the susceptibility peaks should scale $\propto N_s^{\frac{\gamma}{\nu}}$, thus the ratio
$\frac{\gamma}{\nu}$ can be direcly estimated by fitting the peaks according
to $\chi_L^{max}(N_s) = a~N_s^b$.

In Fig.~\ref{fig:1p0GeV2_Nt6_collapseplots} we illustrate the collapse plots
for a first order transition (top panel), for a second order transition of the $\mathbb{Z}_2$
universality class (center panel) and for tricritical scaling (bottom panel).
Data seem to exclude a first order transition and are more compatible
with a continuous transition, even if we do not have enough precision
to discern between the 3D Ising and the tricritical universality class.
A fit of the susceptibility peaks, which is shown in Fig.~\ref{fig:1p0GeV2_Nt6_suscfit},
yields $\frac{\gamma}{\nu} = 2.04(19)$, which also excludes first order and is 
compatible with both 3D Ising and tricritical behavior.
Our conclusion is that for this value of the magnetic field
the transition is still in the $\mathbb{Z}_2$
universality class, but likely not far from a tricritical point separating a first
order region, so that a clear distinction between 3D Ising and tricritial behavior
does not emerge on the explored spatial sizes.

\begin{figure}[t!]
  \centering
  \includegraphics[width=1.0\linewidth, clip]{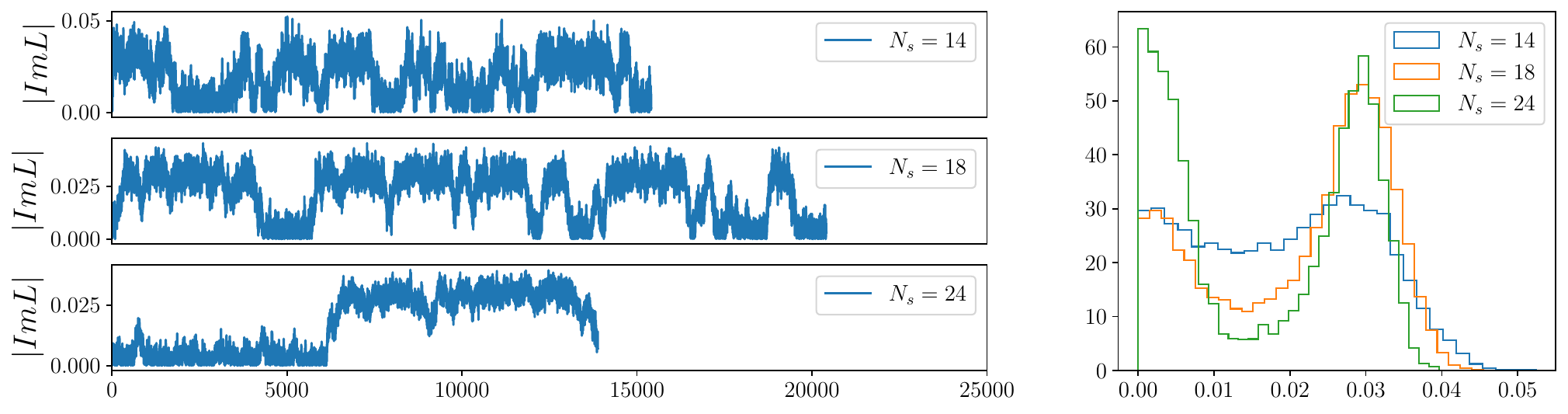}
  \caption{MC histories and histogram: $eB=2.4$ $GeV^2$, $N_t=6$.}
  \label{fig:2p5GeV2_Nt6_hist}
\end{figure}

\begin{figure}[t!]
  \centering
  \includegraphics[width=1.0\linewidth, clip]{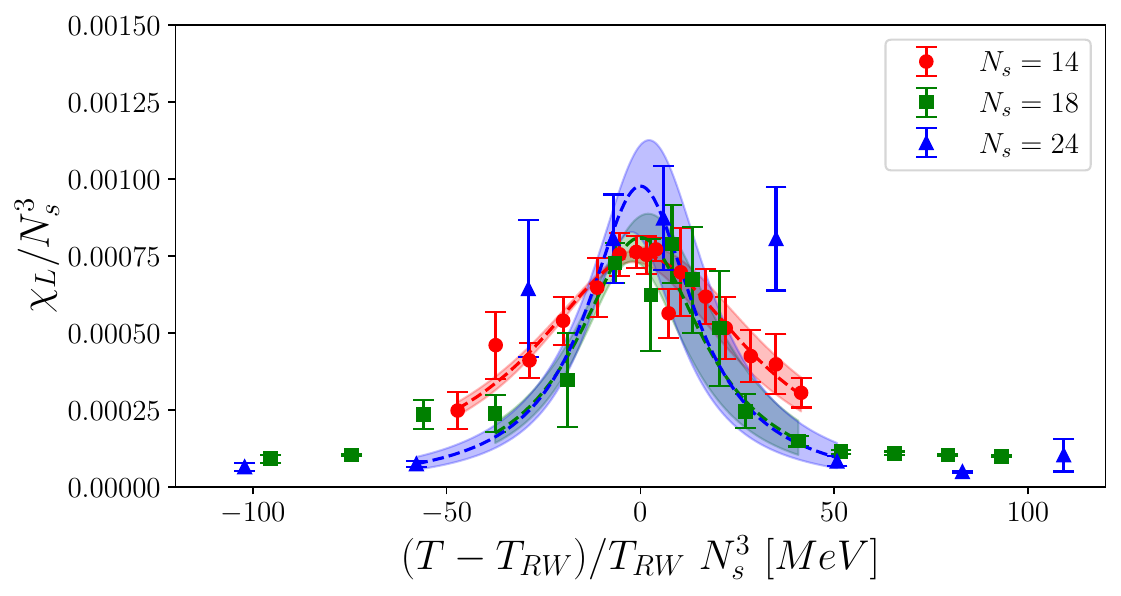}
  \includegraphics[width=1.0\linewidth, clip]{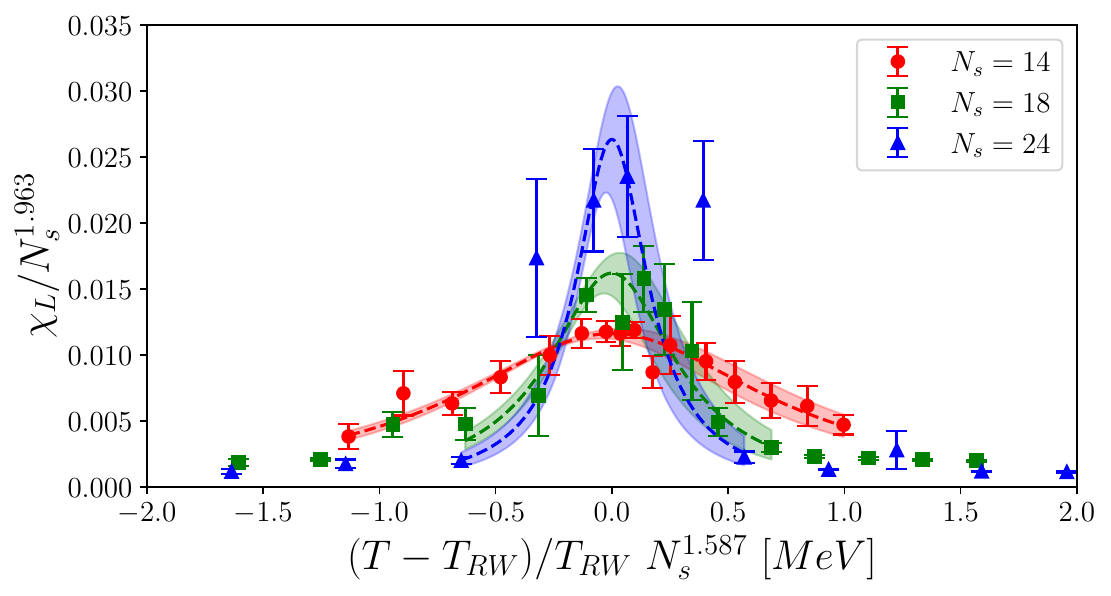}
  \includegraphics[width=1.0\linewidth, clip]{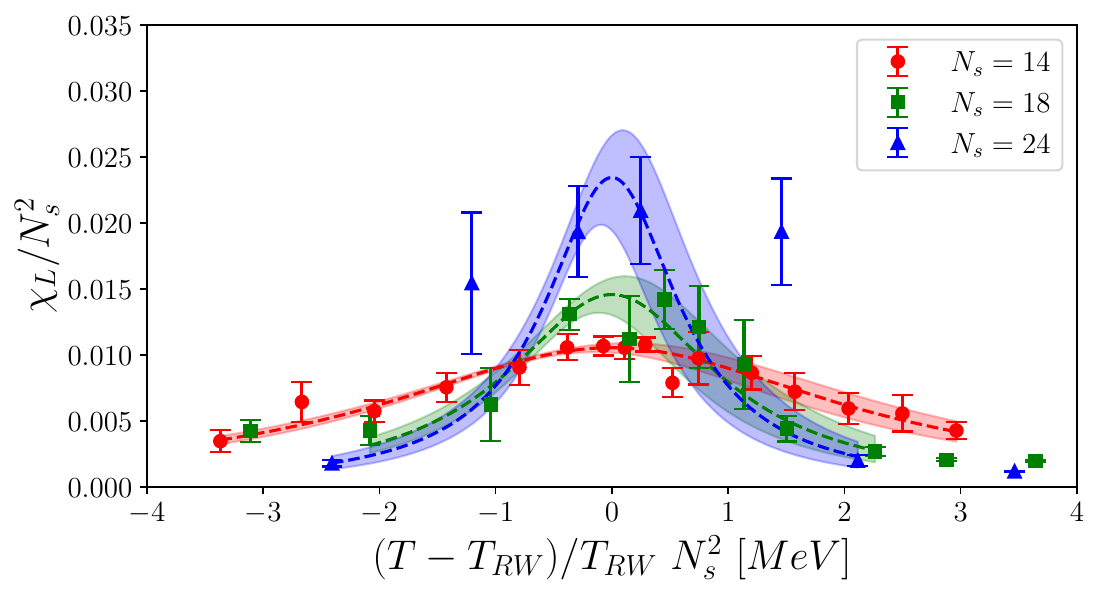}
  \caption{Collapse plots of the order parameter susceptibility for $eB=2.4$ $GeV^2$, $N_t=6$, according to first order (top), second order $Z_2$ (center), tricritical (bottom).}
  \label{fig:2p5GeV2_Nt6_collapseplots}
\end{figure}

\begin{figure}[t!]
  \centering
  \includegraphics[width=1.0\linewidth, clip]{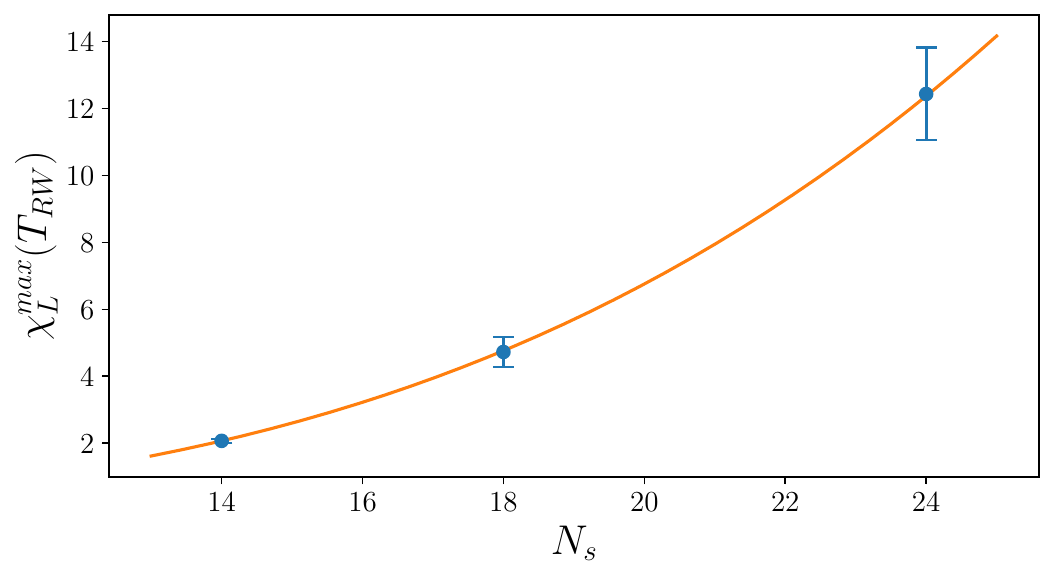}
  \caption{Fit for $\chi_L^{max}$: $eB=2.4$ $GeV^2$, $N_t=6$.}
  \label{fig:2p5GeV2_Nt6_suscfit}
\end{figure}

A qualitatively different picture emerges at $eB = 2.4$~GeV$^2$. This can be seen in
Fig.~\ref{fig:2p5GeV2_Nt6_hist}, which illustrates the Monte Carlo histories (left panel) and histograms
(right panel) of the Polyakov loop obtained close to the critical temperature on lattices with spatial extension
$N_s = 14, 18, 24$. The histograms have a double peaked distribution, which becomes more enhanced as the
volume is increased, suggesting the presence of metastable states typical of a first order phase transition.
The FSS analysis and the best fit to the susceptibility peaks
fully confirms this evidence, as illustrated in Figs.~\ref{fig:2p5GeV2_Nt6_collapseplots}
and Fig.~\ref{fig:2p5GeV2_Nt6_suscfit}. In particular, the best fit
yields $\frac{\gamma}{\nu} = 3.32(20) $, which clearly indicates a first order behavior.

\begin{figure}[t!]
  \centering
  \includegraphics[width=1.0\linewidth, clip]{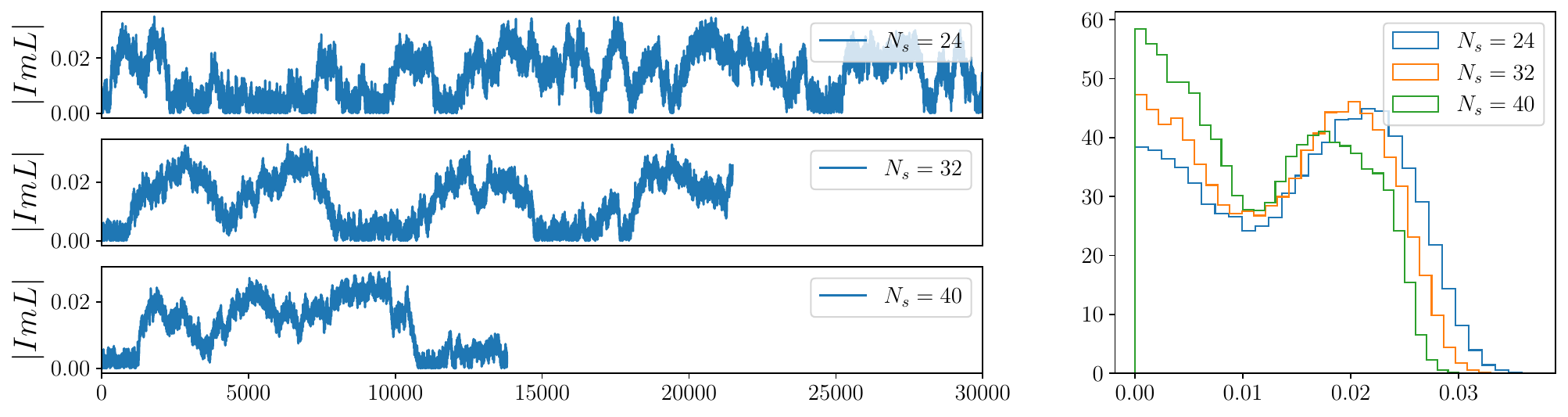}
  \caption{MC histories and histogram: $eB=1.6$ $GeV^2$, $N_t=6$.}
  \label{fig:1p6GeV2_Nt6_hist}
\end{figure}

\begin{figure}[t!]
  \centering
  \includegraphics[width=1.0\linewidth, clip]{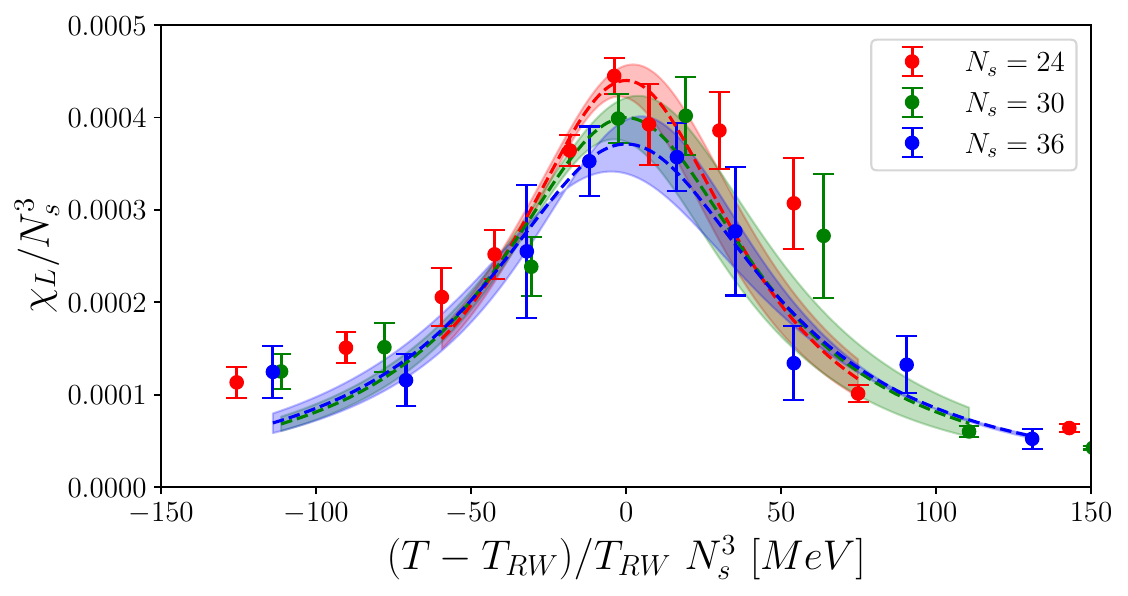}
  \includegraphics[width=1.0\linewidth, clip]{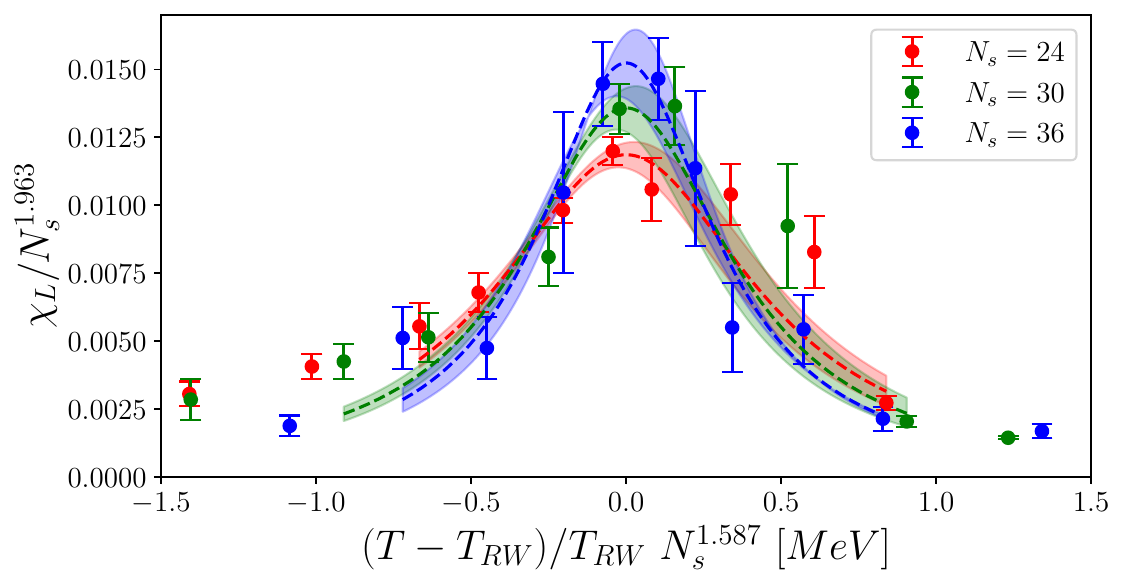}
  \includegraphics[width=1.0\linewidth, clip]{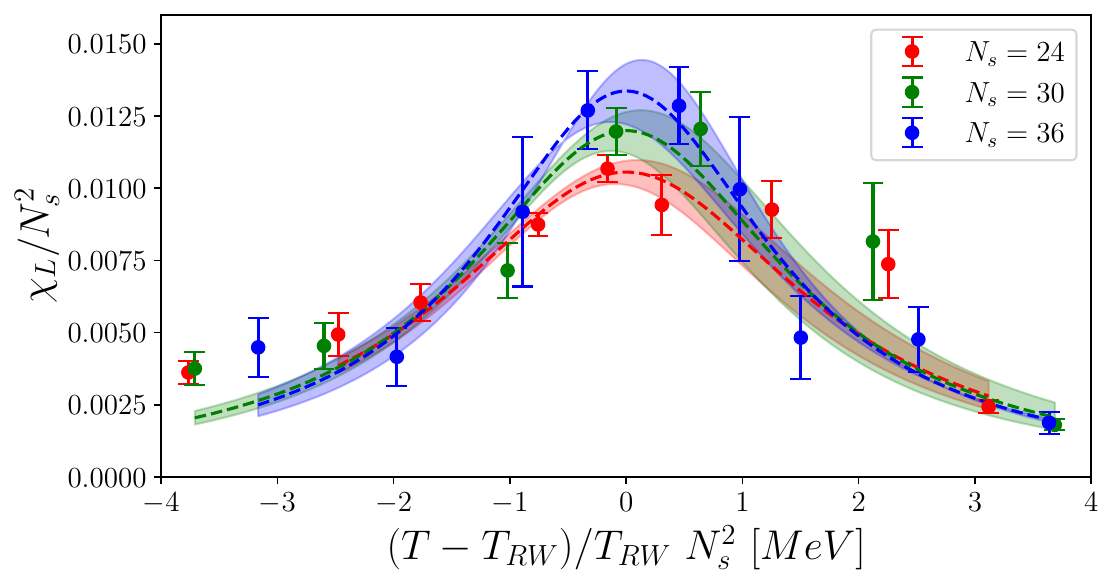}
  \caption{Collapse plots of the order parameter susceptibility for $eB=1.6$ $GeV^2$, $N_t=6$, according to first order (top), second order $Z_2$ (center), tricritical (bottom).}
  \label{fig:1p6GeV2_Nt6_collapseplots}
\end{figure}

\begin{figure}[t!]
  \centering
  \includegraphics[width=1.0\linewidth, clip]{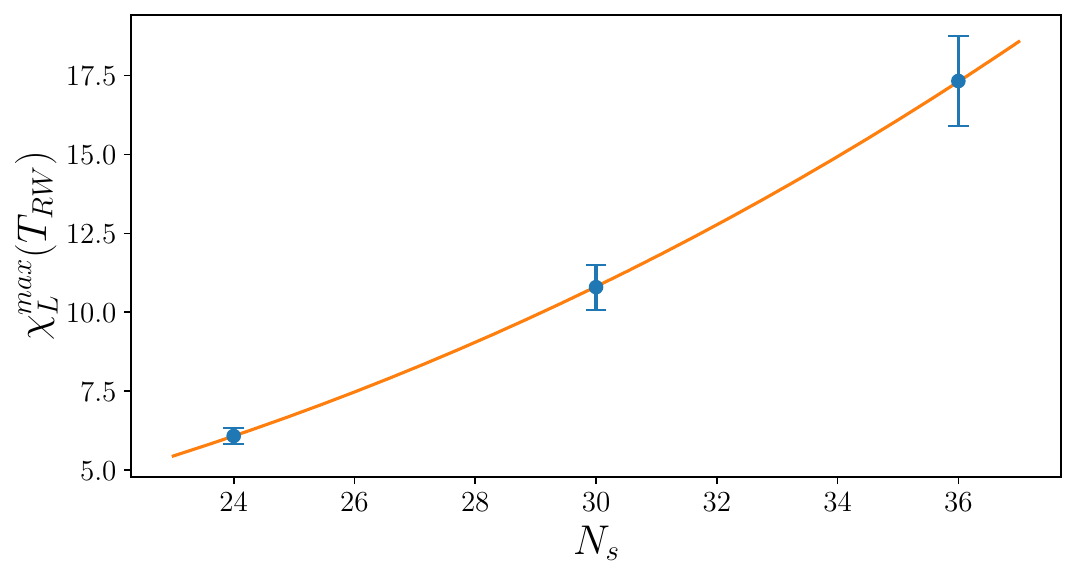}
  \caption{Fit for $\chi_L^{max}$: $eB=1.6$ $GeV^2$, $N_t=6$. The best fit returns
  $\gamma/\nu \sim 2.6$.}
  \label{fig:1p6GeV2_Nt6_suscfit}
\end{figure}

Trying to better locate
the point where the transition turns from second to first order,
we have also performed a FSS analysis for an intermediate value of the magnetic
field, namely $eB = 1.6$~GeV$^2$, on lattices with an extension $N_s = 24, 30, 36$.
Also in this case we illustrate Monte-Carlo histories and
histograms of the order parameter around the transition, see Fig.~\ref{fig:1p6GeV2_Nt6_hist},
a FSS analysis of the order parameter susceptibility
according to different possible critical behaviors, see Fig.~\ref{fig:1p6GeV2_Nt6_collapseplots},
and a best fit to the susceptibility peaks, see Fig.~\ref{fig:1p6GeV2_Nt6_suscfit}.
All these figures show that, despite the increased range of spatial sizes, the analysis
is inconclusive, i.e.~not able to clearly discern the nature of the transition.
On the other hand, this is exactly what one expects close enough to the tri-critical point~\cite{Bonati:2010ce},
since the minimum lattice size needed to clearly discern the correct critical behavior is larger and larger
as the tri-critical point is approached.

Finally for a second intermediate magnetic field $eB = 1.2$ $GeV^2$ we determined the Roberge-Weiss temperature from $N_s=18, 24$ simulations without performing a FSS analysis.

\subsection{Results for $N_t = 8$}

The numerical results obtained on $N_t=6$ lattices clearly indicate that the Roberge-Weiss transition
is still second order at $eB = 1.0$~GeV$^2$, as it is at $eB = 0$,
and first order at $eB = 2.4$~GeV$^2$, thus suggesting the presence
of a tricritical point in-between, which we have not been able to locate more precisely,
based on our present investigation. Given the importance of such indication,
simulations at $eB = 1.0$~GeV$^2$ and $2.4$~GeV$^2$ have been repeated on
$N_t=8$ lattices, in order to confirm the stability of such conclusions as
the continuum limit is approached. This is important especially at $eB = 2.4$~GeV$^2$,
where large discretization effects are possible,
since for $N_t=6$ the magnetic field is not
far from the cut-off $eB = 2\pi / a^2 \sim 5$~GeV$^2$, while this cut-off moves to 9~GeV$^2$
for $N_t = 8$.

\begin{figure}[t!]
  \centering
  \includegraphics[width=1.0\linewidth, clip]{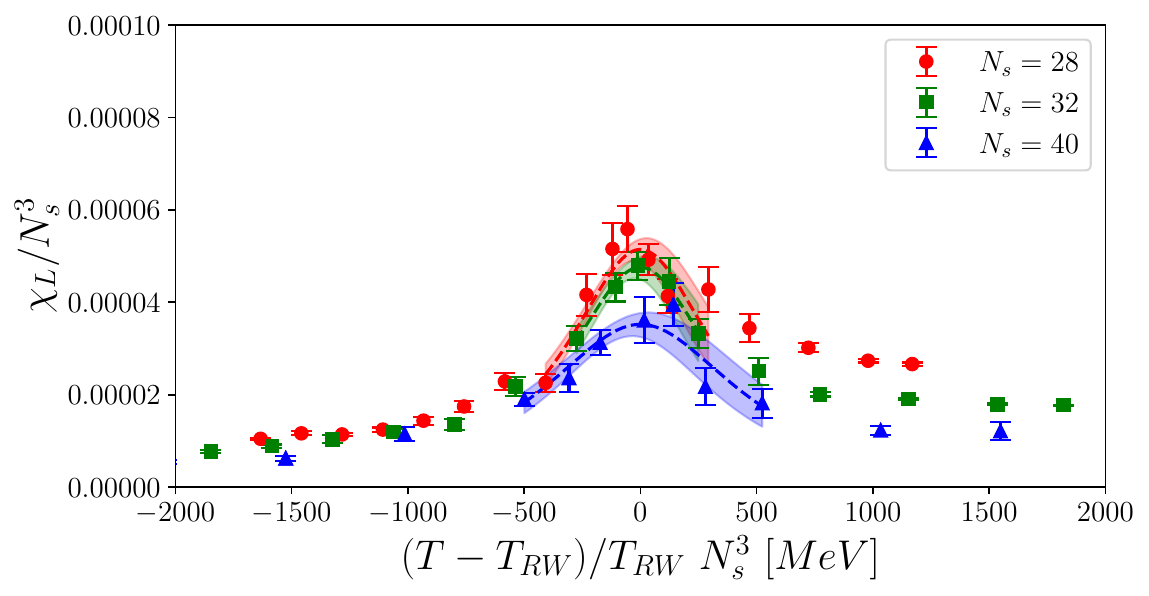}
  \includegraphics[width=1.0\linewidth, clip]{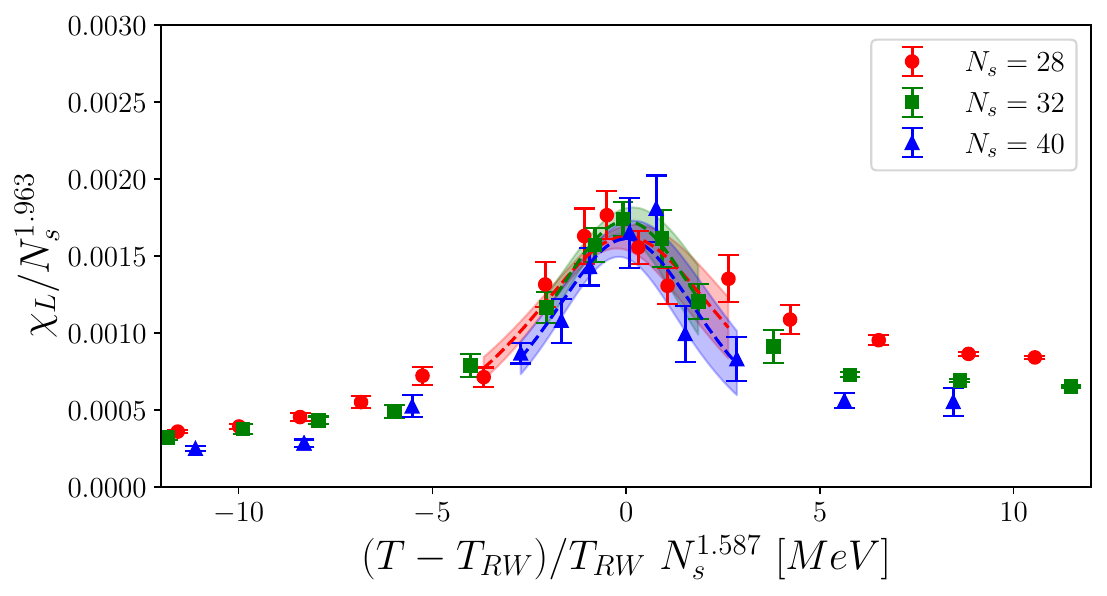}
  \includegraphics[width=1.0\linewidth, clip]{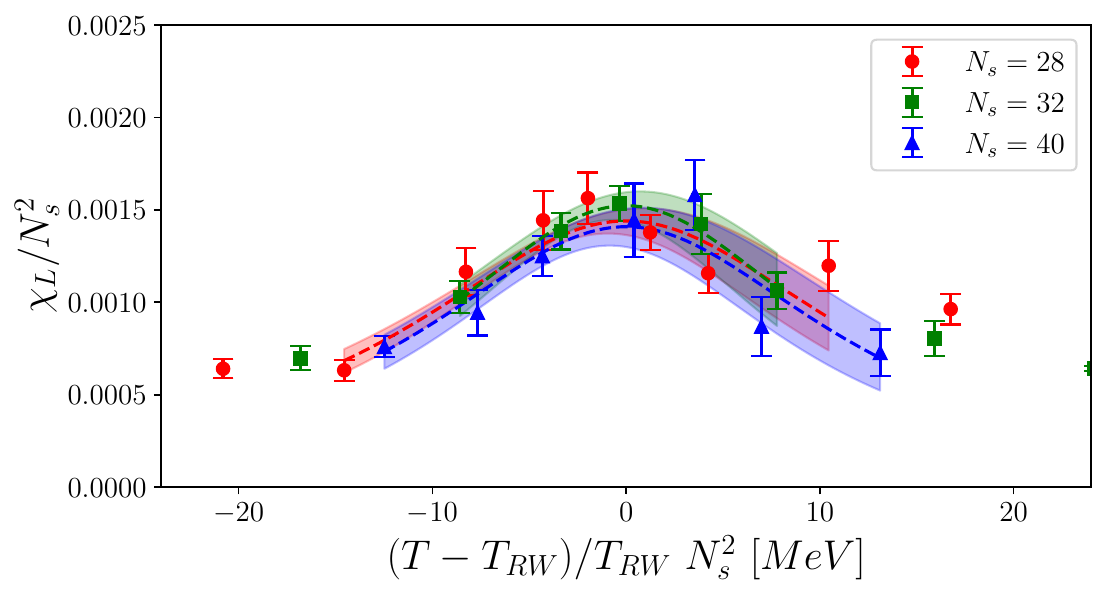}
  \caption{Collapse plots of the order parameter susceptibility for $eB=1.0$ $GeV^2$, $N_t=8$, according to first order (top), second order $Z_2$ (center), tricritical (bottom).}
  \label{fig:1p0GeV2_Nt8_collapseplots}
\end{figure}

\begin{figure}[t!]
  \centering
  \includegraphics[width=1.0\linewidth, clip]{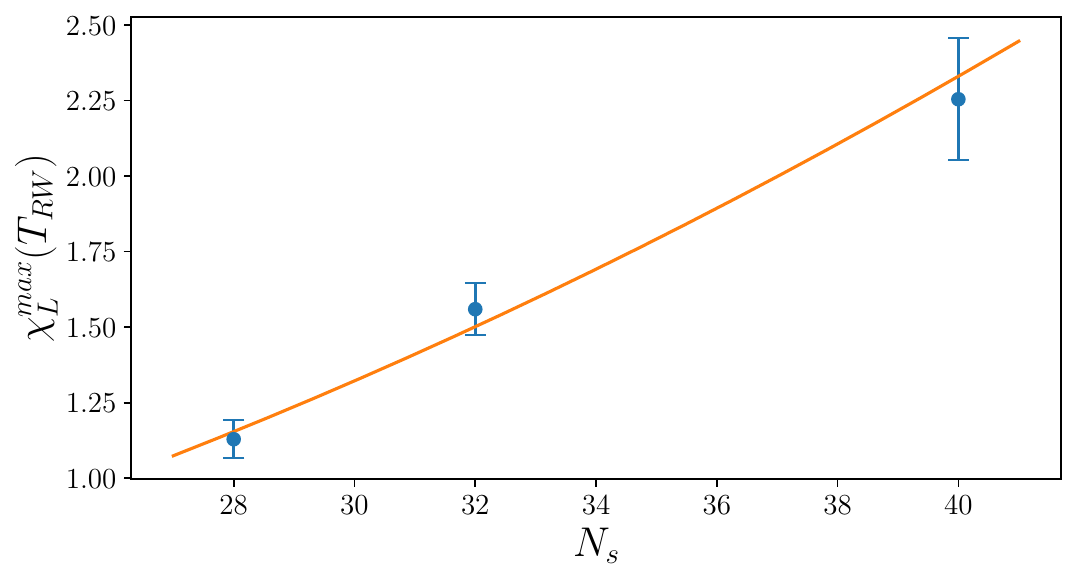}
  \caption{Fit for $\chi_:^{max}$: $eB=1.0$ $GeV^2$, $N_t=8$.}
  \label{fig:1p0GeV2_Nt8_suscfit}
\end{figure}

The results of the FSS analysis at $eB = 1.0$~GeV$^2$ are illustrated in the collapse plots reported
in Fig. \ref{fig:1p0GeV2_Nt8_collapseplots}, indicating again a continuous RW transition even on these
finer lattices. This is further supported by a best fit to the susceptibility peaks,
which yields $\frac{\gamma}{\nu} = 1.97(28)$ (see Fig.~\ref{fig:1p0GeV2_Nt8_suscfit}).

\begin{figure}[t!]
  \centering
  \includegraphics[width=1.0\linewidth, clip]{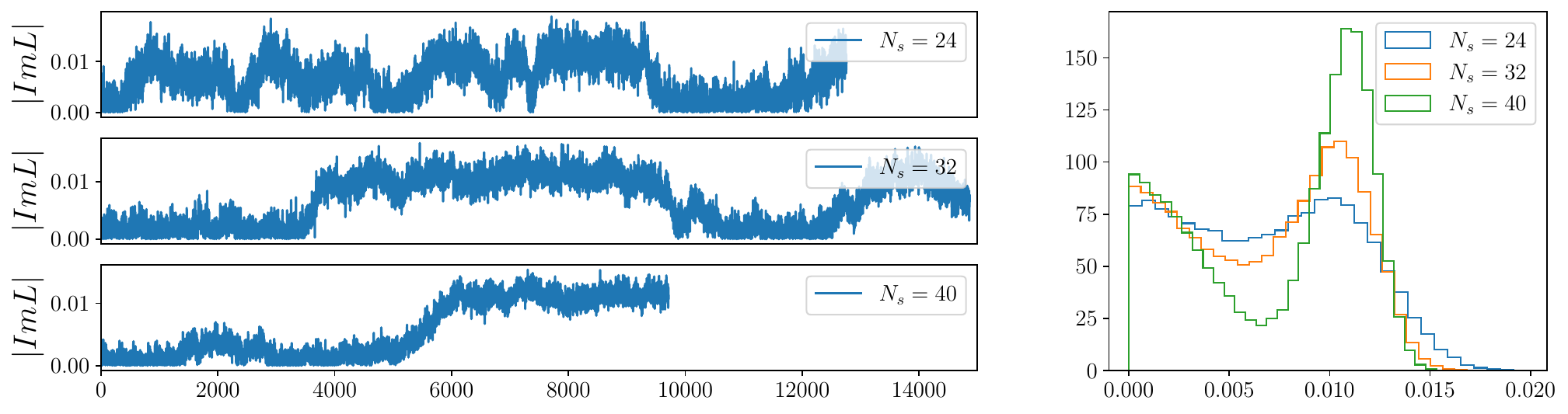}
  \caption{MC histories and histogram: $eB=2.4$ $GeV^2$, $N_t=8$.}
  \label{fig:2p5GeV2_Nt8_hist}
\end{figure}

\begin{figure}[t!]
  \centering
  \includegraphics[width=1.0\linewidth, clip]{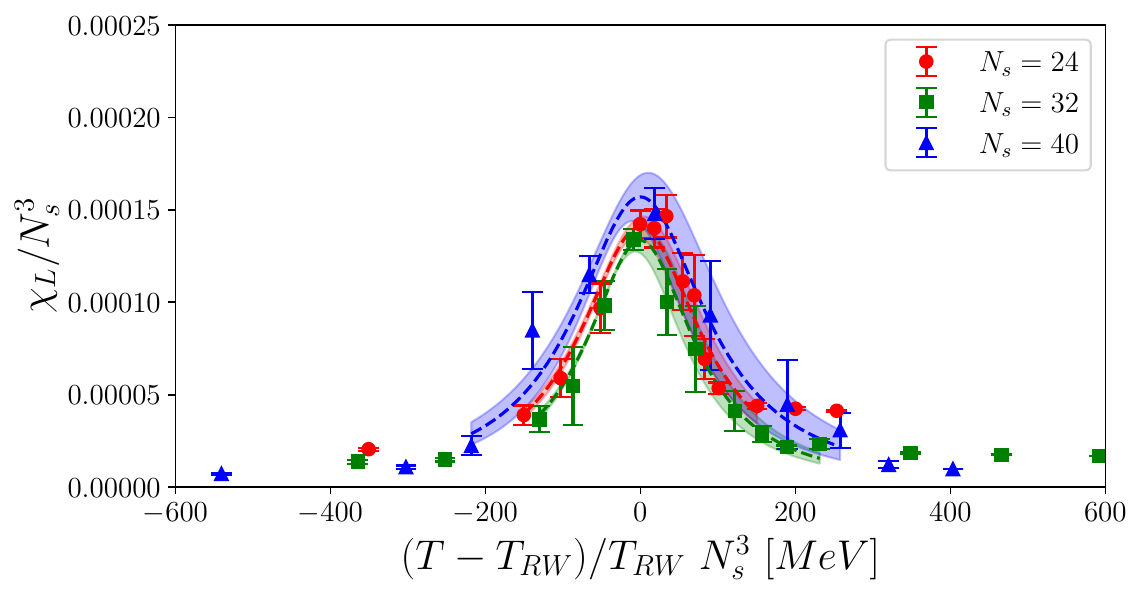}
  \includegraphics[width=1.0\linewidth, clip]{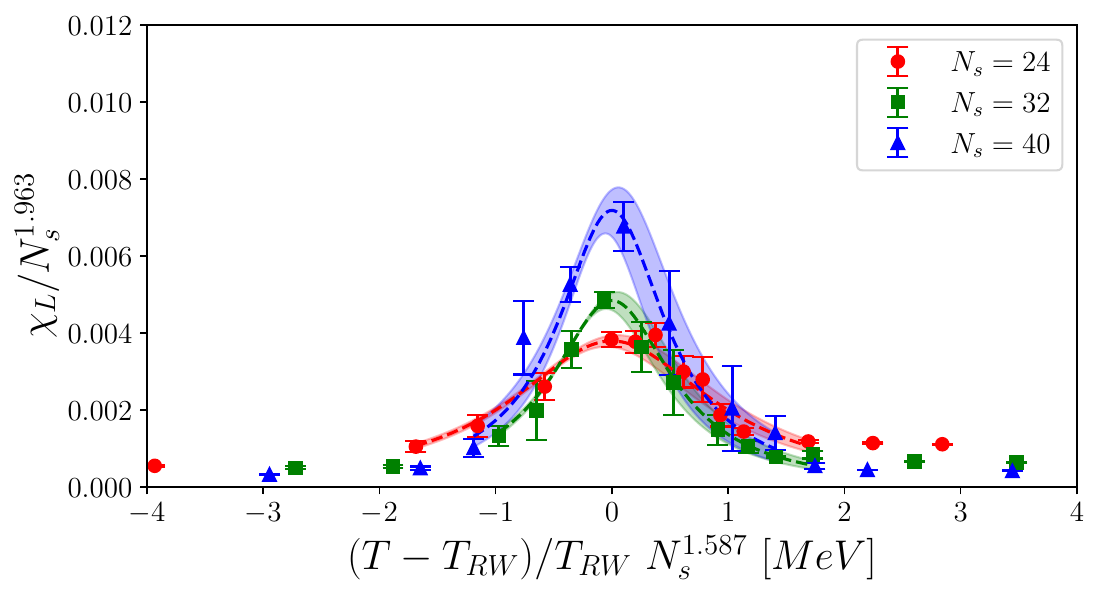}
  \includegraphics[width=1.0\linewidth, clip]{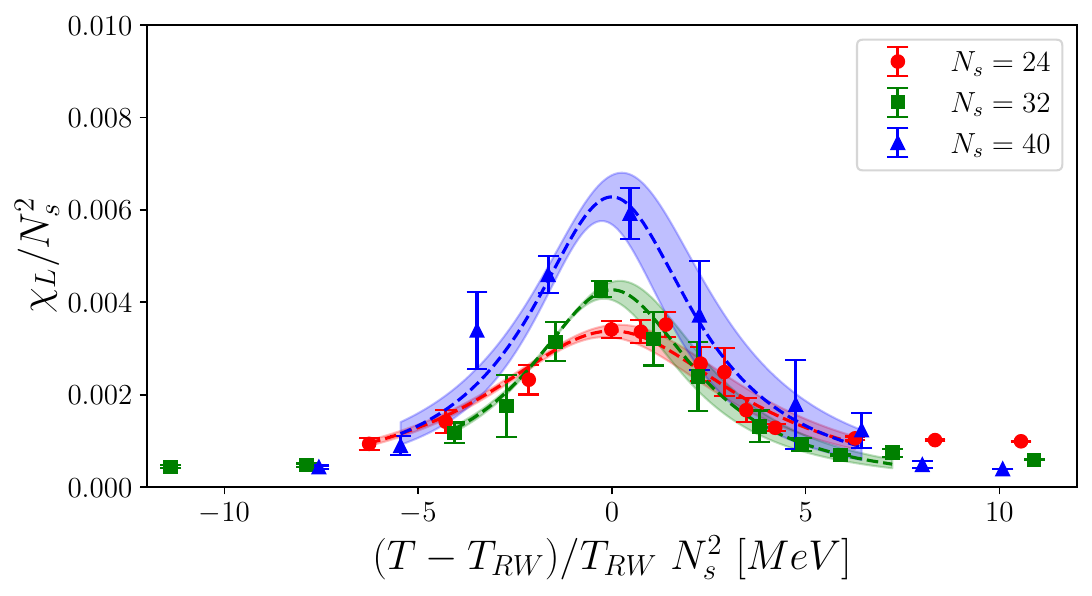}
  \caption{Collapse plots of the order parameter susceptibility for $eB=2.4$ $GeV^2$, $N_t=8$, according to first order (top), second order $Z_2$ (center), tricritical (bottom).}
  \label{fig:2p5GeV2_Nt8_collapseplots}
\end{figure}

\begin{figure}[t!]
  \centering
  \includegraphics[width=1.0\linewidth, clip]{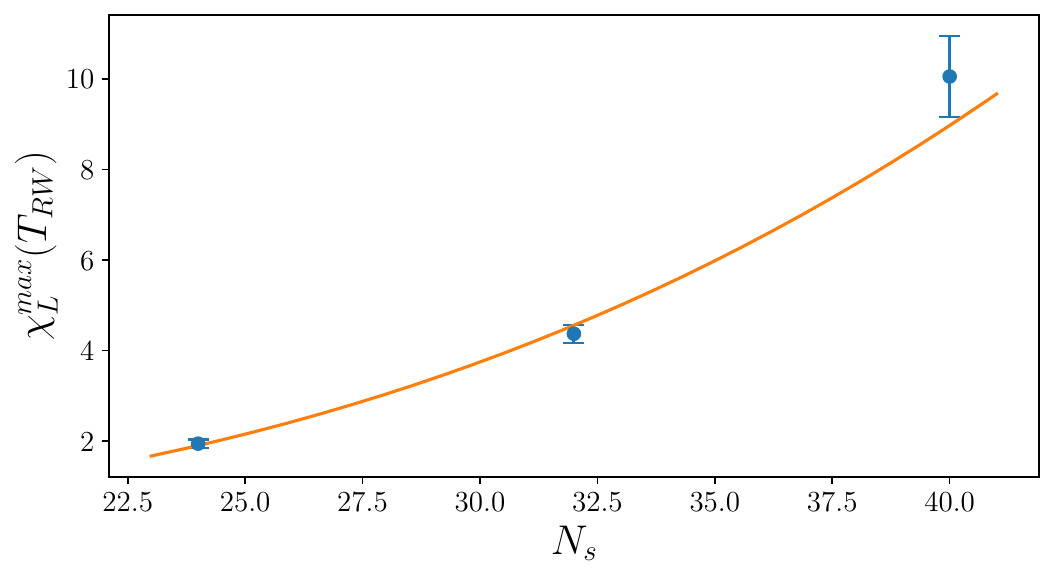}
  \caption{Fit for $\chi_:^{max}$: $eB=2.4$ $GeV^2$, $N_t=8$.}
  \label{fig:2p5GeV2_Nt8_suscfit}
\end{figure}

Conversely, at $eB = 2.4$~GeV$^2$ the Monte Carlo histories and the histograms of the order parameter around
$T_{RW}$, which
are shown respectively in the left and in the right panel of Fig.~\ref{fig:2p5GeV2_Nt8_hist}, still
suggest a clear double peak structure, hence a first order phase transition, which is
confirmed both by the collapse plots displayed in Fig. \ref{fig:2p5GeV2_Nt8_collapseplots}
and by a best fit of the susceptibility peaks, which yields $\frac{\gamma}{\nu} = 3.03(18)$
(see Fig.~\ref{fig:2p5GeV2_Nt8_suscfit}).
We conclude that results obtained on $N_t = 8$ lattices confirm the presence of a
tricritical point in-between 1.0 and 2.4~GeV$^2$.

\subsection{The Roberge-Weiss transition in the $T - eB$ plane}

\begin{figure}[t!]
  \centering
  \includegraphics[width=1.0\linewidth, clip]{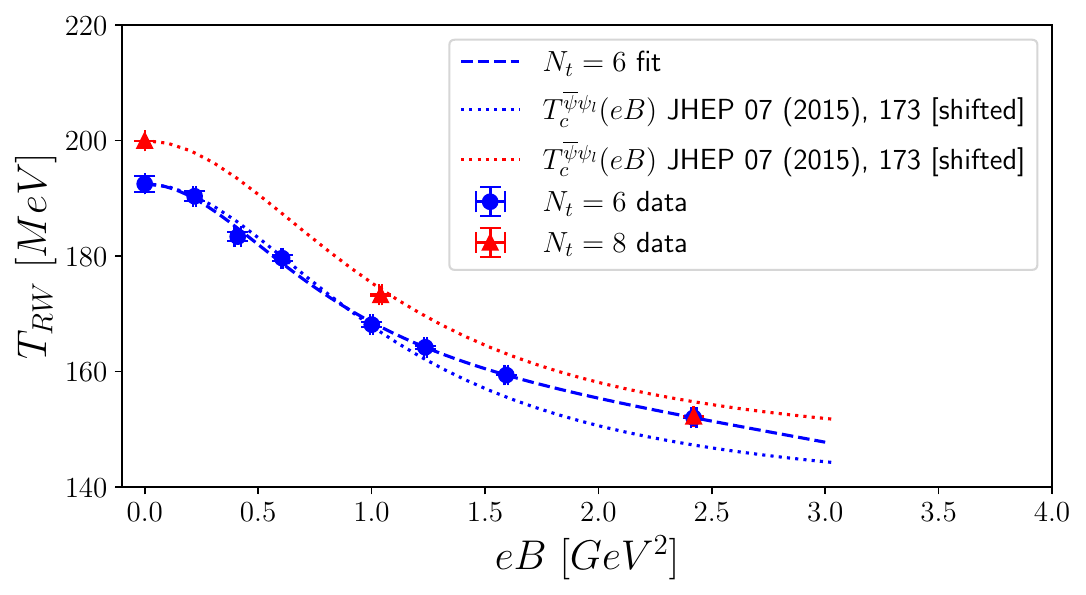}
  \caption{Roberge-Weiss transition line.}
  \label{fig:critline_full}
\end{figure}

So far, we have shown that the Roberge-Weiss temperature decreases monotonically with an increasing magnetic
field, while the strength of the transition increases, so that it becomes first order somewhere
between $1.0$~GeV$^2$ and $2.4$~GeV$^2$. 
Fig.~\ref{fig:critline_full} and Tab.~\ref{tab:critline} summarize the estimates obtained in this work
for the RW transition temperature, both on $N_t = 6$ and $N_t = 8$ lattices.
A rational function ansatz $T^{(2)}_{RW}(eB) = T_{RW}(0) \frac{1 + a(eB)^2}{1 + b(eB)^2}$ fits
well our $N_t=6$ data if the largest explored magnetic field is excluded,
however a higher order rational ansatz is necessary to maintain a
good fit quality when $eB = 2.4$~GeV$^2$ is included.
The dashed blue line represents
a fit done using the rational function ansatz
$T^{(4)}_{RW}(eB) = T_{RW}(0) \frac{1 + a(eB)^2  + c (eB)^4}{1 + b(eB)^2  + d(eB)^4}$.

\begin{table}[bt!]
\begin{tabular}{|c|c|c|}
\hline $N_t$ & $eB$ & $T_{RW}$\\
\hline 
\multirow{5}{*}{$6$} & $0.220(8)$ & $190.37(85)$ \\
& $0.410(14)$ & $183.36(76)$ \\
& $0.606(6)$ & $179.64(83)$ \\
& $1.001(7)$ & $168.01(50)$ \\
& $1.237(7)$ & $164.21(22)$ \\
& $1.594(8)$ & $159.40(5)$ \\
& $2.420(13)$ & $152.02(10)$ \\
\hline
\multirow{2}{*}{$8$} & $1.061(6)$ & 173.30(42) \\
& $2.420(2)$ & 152.24(6) \\ 
\hline
\end{tabular}
\caption{Roberge-Weiss temperatures calculated in this work. Reported errors include only statistical and systematic ones as obtained by means of the fitting procedure, i.e., they do not include systematic errors related to the determination of the lattice scaling, which are common and of the order of 3\% for all data. }\label{tab:critline}
\end{table}

Additionally, in Fig.~\ref{fig:critline_full} we have also reported (as dotted lines)
the rational function parametrization obtained
in Ref.~\cite{Endrodi:2015oba} for the chiral pseudo-critical line in the $(\mu_q = 0, T, eB)$ plane
using the light quark condensate as a probe for the transition.
The parametrization has been shifted along the
temperature axis to match the Roberge-Weiss temperature at zero magnetic field for $N_t=6$ (blue) and $N_t =8$
(red): it is clear that, at least up to $eB = 1$~GeV$^2$, the RW transition line in the
$T - eB$ plane appears as an almost perfect translation of the pseudo-critical chiral transition line.
This is further confirmed by a direct computation of the curvature of the transition line
at $eB = 0$: indeed, by Taylor expanding
our rational function ansatze around $eB = 0$, we find the curvature coefficients $k=-50.0(3.5)$ and
$k=-56(10)$, respectively from the two parametrizations $T^{(2)}_{RW}(eB)$ and $T^{(4)}_{RW}(eB)$.
These values both agree within errors
with the curvature $k \sim -44.8$ which is obtained for the chiral transition line based on
the parametrization found in Ref.~\cite{Endrodi:2015oba}.

\section{Discussion and conclusions}

In this study, we have investigated how a magnetic background field
influences the location and nature of the Roberge-Weiss finite
temperature transition, taking place in the presence of a purely imaginary baryon chemical
potential which is an odd multiple of $\pi$. In particular, we have considered numerical
Monte-Carlo simulations of $N_f = 2+1$ QCD with physical quark masses, discretized through
stout staggered quarks and a tree-level improved Symanzik pure gauge action. Our main findings
can be summarized as follows.

The first interesting observation is that the location of the Roberge-Weiss line as
a function of $eB$,
$T_{RW} (eB)$, appears as an almost perfect (within errors) translation of the
pseudo-critical crossover temperature $T_{pc}(eB)$, i.e., $T_{RW}(eB) - T_{pc}(eB)$
is practically constant, at least up to $eB \sim 1$~GeV$^2$.
Consistent results have been found from the drop of the chiral condensate, which signals chiral
symmetry restoration; in particular the drop of the condensate
moves with $eB$ consistently with what observed for $T_{RW}$, and magnetic catalysis, which
is observed at low $T$, turns into inverse magnetic catalysis around or
above the transition, as a consequence of the drop of the transition temperature.
We consider this as a
further indication of the close similarities between the standard finite temperature
transition and the RW transition, which reflects also on the way the magnetic background
acts on them. Future studies should investigate further this connection, considering
in particular other parameters which are relevant around the transition and refining
the preliminary analysis of chiral symmetry breaking presented in this study.

In addition, also the influence on the nature of the transition is similar, with the magnetic field
which leads to a general strengthening of the transition. In the case of the standard
finite temperature transition, this leads to the appearance of a first order
transition for large magnetic fields, with a critical endpoint located in-between
4 and 9 GeV$^2$~\cite{DElia:2021yvk}. However for the Roberge-Weiss transition, which is second order
at $eB = 0$~\cite{Bonati:2016pwz}, the first order appears much earlier, with a tri-critical
endpoint which we have preliminarily located in-between 1 and 2.4~GeV$^2$.
Although we have not performed a continuum extrapolation, such findings
are stable for two different sets of lattice spacings, corresponding in particular
to $N_t = 6$ and $N_t = 8$; of course, it would be desirable that future studies
refine further the location of such endpoint and its stability when the continuum
limit is actually performed.

\begin{figure}[hbt!]
  \centering
  \includegraphics[width=1.0\linewidth, clip]{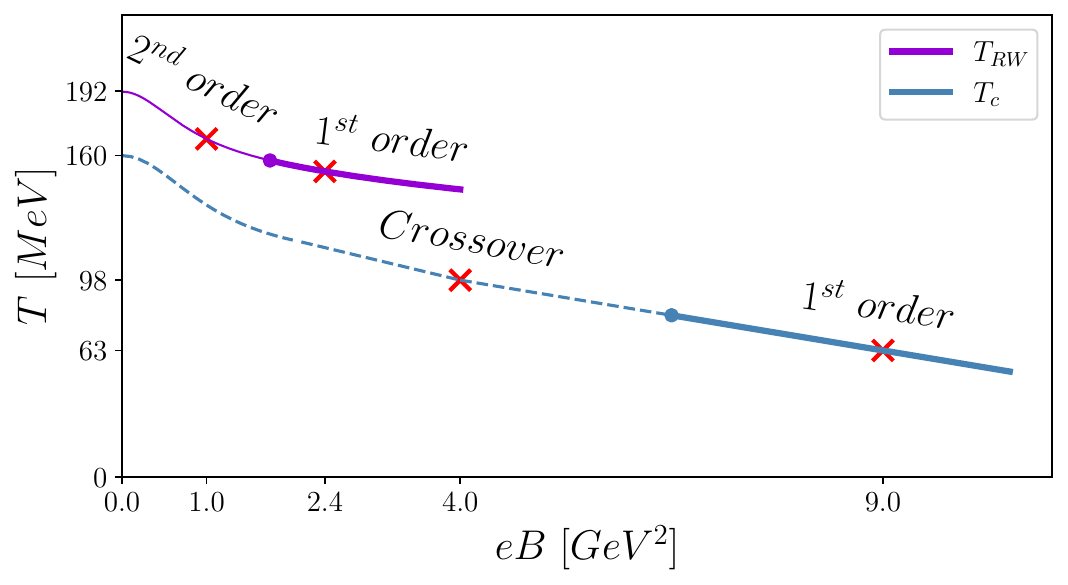}
  \caption{Standard finite temperature transition (lower) and Roberge-Weiss transition (upper) lines
    in the $T - eB$ plane. The location of the critical endpoints should be considered
    only as an approximate and rough interpolation of the findings obtained
    by numerical simulations, performed at the points denoted by red crosses.}
  \label{fig:critical_TcTRW2}
\end{figure}

\begin{figure}[hbt!]
  \centering
  \includegraphics*[width=1.0\linewidth, clip]{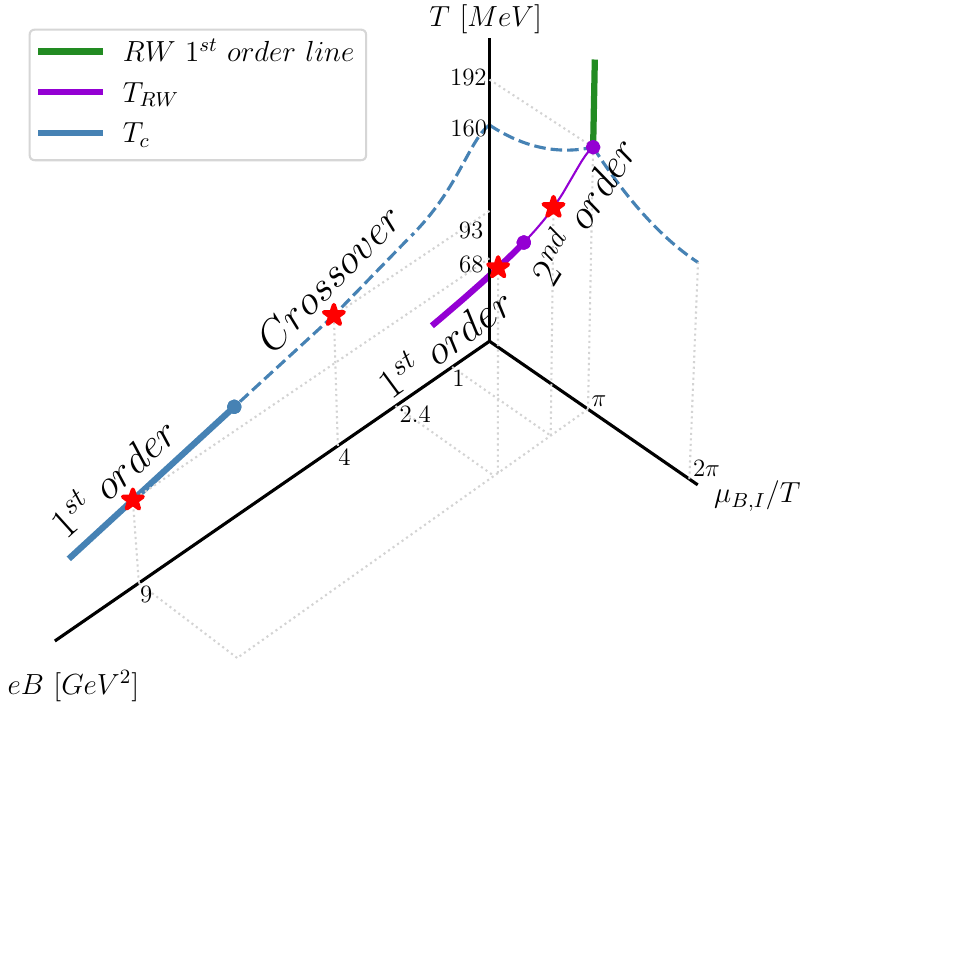}
  \vspace{-2.5cm} 
  \caption{Extended phase diagram in the $T - \mu_{B,I}/T - eB$ space, based on the findings of this investigations
    and of previous studies. One should imagine an extension of the analytic continuation of the pseudo-critical line
    in the $T - \mu_{B,I}$ plane, which connects the standard QCD crossover to the Roberge-Weiss transition, in order
    to form transition surfaces in this extended space.}
  \label{fig:critical_TcTRW3}
\end{figure}

We illustrate a summary of our findings in Figure~\ref{fig:critical_TcTRW2}, where both
the standard finite temperature transition lines and the RW transition lines are drawn
in the $T - eB$ plane, according to the results of this and of previous investigations.
In order to get  a better perspective on the QCD phase diagram as a whole, we report
the same findings in Figure~\ref{fig:critical_TcTRW3}
in  the three-dimensional $T - \mu_{B,I}/T - eB$ space, where we have also
added the available information on the phase diagram which is found
for imaginary baryon chemical potentials at $eB = 0$.

Such information at $eB = 0$ consists basically in 
the presence of a high-$T$ first order line (Roberge-Weiss line) with an endpoint (Roberge-Weiss
finite temperature transition) which connects to the standard chiral crossover temperature through
the analytic continuation of the QCD pseudo-critical line; that repeats periodically
in $\mu_{B,I}$, according to the expected Roberge-Weiss periodicity. One should imagine,
looking at Figure~\ref{fig:critical_TcTRW3}, that such structure is continued to non-zero
$eB$, in order to form pseudo-critical or critical surfaces in the extended phase diagram. 

Within this view, the critical and tri-critical endpoints,
which are found respectively at zero chemical potential and at the RW point, should be connected
by a line of critical endpoints for intermediate values of the imaginary chemical potential,
which could be better defined by future studies. Apart from such refinements, this already
gives a strong indication that the effect of adding an imaginary chemical potential
is to move the critical endpoint in the $T - eB$ plane towards lower values of $eB$.

This seems
at odd with the presence of a strict connection between the $T - eB$ and the $T - \mu_B$ phase diagrams,
despite the close similarities, consisting in particular in
the presence of a first order transition at low temperature with an associated endpoint.
Indeed, as we have discussed in the Introduction, in this case one would expect that the critical magnetic field decreases as a function
of $\mu_B$, hence that it increases as a function of ${\rm Im} (\mu_B) / T$, assuming
the theory is analytic around $\mu_B = 0$ and that first contributions are quadratic in $\mu_B$.

\label{sec:conclusions}

\acknowledgements

This work has also been partially supported
by the project “Non-perturbative aspects of fundamental interactions, in the Standard Model and beyond” funded by MUR,
Progetti di Ricerca di Rilevante Interesse Nazionale (PRIN), Bando 2022, grant 2022TJFCYB (CUP I53D23001440006).
Numerical simulations have been performed on the Marconi 100 and Leonardo clusters at CINECA, based on the
agreement between INFN and CINECA under projects INF23\_npqcd and INF24\_npqcd. KZ acknowledges support by the
project “Non-perturbative aspects of fundamental interactions, in the Standard Model and beyond” funded by MUR,
Progetti di Ricerca di Rilevante Interesse Nazionale (PRIN), Bando 2022, Grant 2022TJFCYB (CUP I53D23001440006).

\bibliography{biblio}

\end{document}